\documentclass[twocolumn]{aastex62}
\usepackage{amsmath}
\usepackage{multirow}
\usepackage{breakurl}
\usepackage{hyperref}
\usepackage{textcomp}
\usepackage{cleveref}[2012/02/15]
\usepackage{fancyhdr}
\usepackage{graphicx}
\usepackage{amssymb}
\usepackage[toc,page]{appendix}

\crefformat{footnote}{#2\footnotemark[#1]#3}

\newcommand{\HI}{\ion{H}{1}}

\newcommand{\kms}{km~s$^{-1}$}
\newcommand{\s}{$\sim$}
\newcommand{\n}{$-$}

\newcommand{\soobshchspets}{Astrof. Issledovanija Byu. Spec. Ast. Obs.}

\shortauthors{Ashley, T. et al.}

\begin{document}

\title{The Neutral Gas Properties of Extremely Isolated Early-Type Galaxies III}

\author{Trisha Ashley}
\affiliation{Space Telescope Science Institute}
\affiliation{NASA Ames Research Center}
\email{tashley@stsci.edu}

\author{Pamela M. Marcum}
\affiliation{NASA Ames Research Center} 
\email{pamela.m.marcum@nasa.gov}

\author{Mehmet Alpaslan}
\affiliation{New York University} 
\email{ma5046@nyu.edu}

\author{Michael N. Fanelli}
\affiliation{NASA Ames Research Center} 
\email{michael.n.fanelli@nasa.gov}

\author{James D. Frost}
\email{jim.frost@gmail.com}

 \begin{abstract}
We report on the neutral hydrogen gas content (21-cm emission) of eight extremely isolated early-type galaxies (IEGs) using the Green Bank Telescope. Emission is detected in seven of the eight objects. This paper is the third in a series that collectively present new \HI\ observations for 20 IEGs. Among the 14 \HI\ detections in our observations, eight exhibit a Gaussian-like \HI\ line profile shape, four are double-peaked, one is triple-peaked, and another has a plateaued rectangular shape.  Five additional IEGs observed in previous surveys were added to our analysis, bringing the total number of IEGs with \HI\ observations to 25. Of these objects, emission is detected in 19 (76\%). The 25 IEGs in our combined study have gas masses that are systematically larger than their luminosity-matched comparison galaxy counterparts. The isolated early-type galaxies presented here follow a trend of increasing gas-richness with bluer B\n V colors. This correlation is also observed in a comparison sample drawn from the literature composed of loose group and field early-type galaxies. Two IEGs, KIG164 and KIG870, exhibit properties highly anomalous for spheroidal systems: luminous (M$_{\rm B}$ = \n20.5, \n20.1) and blue (B\n V = 0.47, 0.48) respectively, with substantial neutral gas, M$_{\rm{HI}}$ = 4.1 \& 5.5 $\times 10^9$ M$_{\sun}$. Other IEG systems may represent early-type galaxies continuing to assemble via quiescent \HI\ accretion from the cosmic web or relaxed merged systems.

\end{abstract} 
 
\keywords{galaxies: elliptical and lenticular, cD -- galaxies: evolution -- galaxies: ISM -- galaxies: star formation }

\section{Introduction}\label{sect:intro}

Early-type galaxy (ETG) formation and evolution is likely to be driven by multiple mergers, a process facilitated in cluster environments in which ETGs preferentially reside. Isolated early-type galaxies (IEGs) present a unique opportunity to test this formation model and to study the nature of ETG evolution outside of the influences of the massive cluster environment e.g., ram pressure stripping, ionizing intracluster material, tidal harassment, and gas strangulation \citep{Larson1980apj237_692, Moore1999mnras304_465, Gunn1972apj176_1, Lokas2016apj826_227}. The isolated nature of these systems strongly limits the possible sequences of events leading to their formation with primary mechanisms including early-epoch monolithic collapse or the coalescence of an isolated group of galaxies \citep{Marcum2004aj127_3213, Niemi2010mnras405_477, Naab2007apj658_710}.

ETGs in low density environments are notably bluer and more gas-rich than their cluster ETG counterparts \citep{Huchtmeier1995aap300_675, Aars2001aj122_2923, Marcum2004aj127_3213, Niemi2010mnras405_477, Fuse2012aj144_57, Ashley2017aj153_158, Ashley2018aj155_15}. The bluer colors are suggestive of higher star formation rates over the past $\sim0.5$~Gyr \citep{Bruzual2003mnras344_1000}. Cold gas reservoirs that can be used to fuel star formation in such ETGs are likely to remain unperturbed outside of the dynamically active cluster environment.  Potential origins of gas reservoirs include accretion of surrounding gas clouds and/or recent mergers of nearby companions \citep{Keres2005mnras363_2, Maccio2006mnras366_1529}. Some isolated ETGs may also still be in the process of forming and have not had sufficient time to ionize, expel, and/or consume a majority of their neutral gas \citep{VanDriel1991aap243_71, Barnes2002mnras333_481, Niemi2010mnras405_477, Serra2012mnras422_1835}. While the \HI\ content of field ETGs and correlations with local environment has received increased attention (e.g.  \citealt{Morganti2006mnras371_157}, \citealt{Oosterloo2010mnras409_500},
\citealt{Serra2012mnras422_1835}), significantly fewer \HI\ studies of \textit{isolated} or void ETGs have been conducted (three in \citet{Kreckel2012aj144_16}, 12 in \citealt{Ashley2017aj153_158, Ashley2018aj155_15}). The Void Galaxy Survey \citep{Kreckel2012aj144_16} measured the \HI\ content and kinematics for 60 morphologically diverse systems residing in voids identified in the Sloan Digital Sky Survey, detecting 41 generally \HI\ gas-rich, low-luminosity, blue disk galaxies. None of the three morphologically classified ETGs in their survey was detected in \HI.  Based on the observed gas kinematics and star formation activity, \citet{Kreckel2012aj144_16} suggest that the void galaxy population is still in the process of assembling. A number of the Void Galaxy Survey objects have nearby companions and are therefore not necessarily locally isolated system. The highly isolated galaxies presented in this paper allow a study of galaxy gas content to local galaxy number densities even lower than those of general void systems.  

\HI\ spectra provide insight into both the kinematics of neutral gas and the fuel available for star formation. These data facilitate understanding of the origin, size, and longevity of the \HI\ reservoirs in highly isolated systems and permit comparison to galaxies evolving in higher density environments.  This paper is the third in a series discussing the \HI\ content of extremely isolated early-type galaxies.  Two previous papers, \citet{Ashley2017aj153_158, Ashley2018aj155_15}, hereafter Papers 1 \& 2, presented Green Bank Telescope (GBT) \HI\ data for 12 IEGs.  This paper presents GBT \HI\ data for eight additional IEGs, accounting for all 20 objects observed by us. Five additional galaxies for which \HI\ data was published by other groups is included in the analysis. This study of 25 isolated objects increases both the number of such systems observed in \HI\ as well as the range in properties of the sample, e.g., luminosity and color, for which \HI\ data exist. Sections~2 discusses the IEG and comparison samples. Section~3 presents the observations, data reduction, and measured and derived \HI\ properties.  Section~4 reviews the individual \HI\ profiles and compares gas-richness of the IEGs to the comparison sample.  Section~5 summarizes the results.

\section{Sample Selection}\label{sect:sampleselect}

\subsection{Isolated Early-type Galaxy Sample}\label{sect:iegsample}
The sample of extremely isolated early-type galaxies presented here is drawn from the compilations of \citet{Marcum2004aj127_3213} and \citet{Fuse2012aj144_57} (M04 \& F12). An ``early-type'' morphology was assigned to galaxies with bulge-to-total light ratios (B/T) of $\ge$ 0.6 \citep{Marcum2004aj127_3213} or inverse concentration index ratios (R$_{50}$/R$_{90}$) of $\le$0.38 \citep{Fuse2012aj144_57}, resulting in a dataset including both elliptical and S0 galaxies ($-5 < T \le -1$). The parent samples were chosen to be separated by at least 2.5~Mpc in co-moving distance from neighboring galaxies brighter than M$_{\text V}$ $=$ \n16.5 (\s2 magnitudes dimmer than ATLAS 3D), eliminating the possibility of past interactions with any non-dwarf companions \citep{Aars2001aj122_2923}.  Using deep optical images, M04 \& F12 found evidence of past interactions and mergers in several of the IEGs, including asymmetric outer isophotes and multiple nuclei.  The M04 sample was drawn from the Catalog of Isolated Galaxies \citep{Karachentseva1973soobshchspets8_3} and contains nine IEGs brighter than M$_{\text B} < -19.6$ with a range of colors, 0.46 $ < $ B\n V $ < 0.96$. The larger F12 sample, extracted from the Sloan Survey, is on average bluer in color, less luminous, and smaller in physical extent than the M04 set, with  average values of B\n V$=0.63$, $L_B =1.13\times 10^9$ M$_{\sun}$, and optical radius of \s4.5 kpc. Consistent with the approach adopted for the \HI\ photometry presented in Papers~1 \& 2, galaxies predicted to have an \HI\ line flux sufficient to provide a 5$\sigma$ detection within a $\sim$10-hour exposure were selected as targets.  Of the 35 IEGs considered for GBT observations presented here and in Papers~1 \& 2, twelve objects were eliminated as viable GBT targets due to large exposure times estimated for a threshold detection.

\begin{deluxetable*}{lcrrcccclcr}
\tablecaption{Basic Galaxy Information\label{tab:galinfo}}
\tabletypesize{\footnotesize}
\tablecolumns{11}
\tablewidth{0pt}
\tablehead{
\colhead{Name} & \colhead{Abbrev. Name} & \colhead{RA (2000)} &  \colhead{Dec (2000)} &  \colhead{Distance} & \colhead{Size} & \colhead{Size} & \colhead{V} & \colhead{M$_{\rm{B}}$} & \colhead{B$-$V} & \colhead{L$_{\rm{B}}$}\\ 
\colhead{} & \colhead{} & \colhead{(hh mm ss.s)} & \colhead{(dd mm ss)} & \colhead{(Mpc)} & \colhead{(\arcsec)} & \colhead{kpc} & \colhead{(\kms)} & \colhead{(mag)} &  \colhead{(mag)} & \colhead{(10$^8$ L$_{\sun}$)}\\
\colhead{(1)}  & \colhead{(2)} & \colhead{(3)}  & \colhead{(4)} & \colhead{(5)} & \colhead{(6)}  & \colhead{(7)}  & \colhead{(8)} & \colhead{(9)} & \colhead{(10)} & \colhead{11}}
\startdata
KIG164 & K164 & 05 55 03.68 & +74 42 52.88 & 127 & 33 & 20.3 & 9033 & \n20.5 & 0.47 & 234.1\\
SDSSJ092657.99+100301.4 & S0926 & 09 26 57.98 & +10 03 01.40 & 73 & 28 & 9.9 & 5313 & \n18.0\tablenotemark{a} & 0.24 & 24.3\\
SDSSJ094410.80+001047.3 & S0944 & 09 44 10.90 & +00 10 47.11 & 46 & 26 & 5.8 & 3333 & \n16.5\tablenotemark{a} & 0.47 & 5.9\\
SDSSJ104807.06+430525.5 & S1048 & 10 48 07.05 & +43 05 25.47 & 58 & 17 & 4.8 & 4005 & \n15.8\tablenotemark{a} & 0.41 & 3.1\\
SDSSJ111029.62+134558.1 & S1110 & 11 10 29.61 & +13 45 58.12 & 61 & 22 & 6.5 & 4331 & \n16.8\tablenotemark{a} & 0.44 & 7.5\\
KIG684 & K684 & 15 27 14.86 & +77 09 24.61 & 78 & 50 & 19.0 & 5380 & \n19.5 & 0.98 & 91.8\\
SDSSJ155325.18+520416.5 & S1553 & 15 53 25.18 & +52 04 16.50 & 48 & 28 & 6.5 & 3109 & \n16.9\tablenotemark{a} & 0.47 & 8.8\\
SDSSJ212753.52\n070113.9 & S2127 & 21 27 53.52 & \n07 01 13.92 & 49 & 38 & 9.1 & 3519 & \n17.9\tablenotemark{a} & 0.36 & 20.8\\
\cutinhead{Basic Information for IEGs Presented in Other Papers}
SDSSJ024248.69\n082356.6 & S0242 & 02 42 48.69 & \n08 23 56.63 & 58 & 38 & 10.7 & 4337 & \n17.5\tablenotemark{a} & 1.03 & 14.4\\
NGC1211 & N1211 & 03 06 52.41 & \n00 47 40.14 & 43 & 92 & 19.1 & 3216 & \n20.2 & 0.86 & 174.5\\
VIIIZw40 & Z40 & 09 11 05.57 & +09 20 58.40 & 51 & 26 & 6.3 & 3660 & \n17.4\tablenotemark{a} & 0.90 & 13.2\\
SDSSJ102145.89+383249.8 & S1021 & 10 21 45.89 & +38 32 49.78 & 60 & 18 & 5.3 & 4204 & \n17.1\tablenotemark{a} & 0.34 & 10.5\\
Mrk150 & M150 & 10 38 37.24 & +44 31 23.08 & 54 & 49 & 12.9 & 3700 & \n18.3 & 0.35 & 31.4\\
SDSSJ113237.43+472658.7 & S1132 & 11 32 37.43 & +47 26 58.67 & 25 & 21 & 2.6 & 1453 & \n14.2\tablenotemark{a} & 0.50 & 0.7\\
Mrk737 & M737 & 11 35 23.86 & +31 39 15.16 & 49 & 31 & 7.3 & 3348 & \n17.2\tablenotemark{a} & 0.28 & 11.8\\
SDSSJ122123.13+393659.5 & S1221 & 12 21 23.12 & +39 36 59.48 & 65 & 13 & 4.2 & 4482 & \n15.9\tablenotemark{a} & 0.36 & 3.3\\
KIG557 & K557 & 12 55 16.57 & +00 14 48.81 & 197 & 39 & 37.0 & 14304 & \n20.7 & 1.02 & 274.2\\
SDSSJ132337.69+291717.1 & S1323 & 13 23 37.69 & +29 17 17.08 & 59 & 16 & 4.6 & 4068 & \n15.6\tablenotemark{b} & 0.49 & 2.6\\
SBS1327+597 & S1327 & 13 27 18.56 & +59 30 10.25 & 72 & 32 & 11.3 & 4956 & \n18.0\tablenotemark{a} & 0.92 & 24.2\\
SDSSJ143538.00+435937.1 & S1435 & 14 35 38.00 & +43 59 37.10 & 77 & 20 & 7.5 & 5333 & \n17.0\tablenotemark{a} & 0.46 & 9.3\\
KIG792 & K792 & 17 11 51.43 & +41 38 57.69 & 129 & 47 & 29.6 & 9188 & \n20.4 & 0.89 & 209.3\\
KIG824 & K824 & 17 40 38.03 & +41 18 04.62 & 78 & 54 & 20.4 & 5352 & \n19.5 & 0.89 & 98.4\\
KIG870 & K870 & 19 11 48.90 & +83 54 20.27 & 89 & 50 & 21.7 & 6197 & \n20.1 & 0.48 & 160.3\\
SDSSJ234042.69\n092336.6 & S2340 & 23 40 42.68 & \n09 23 36.59 & 70 & 30 & 10.1 & 5146 & \n18.2\tablenotemark{a} & 0.45 & 29.1\\
SDSSJ235021.45+141342.6 & S2350 & 23 50 21.45 & +14 13 42.60 & 72 & 16 & 5.4 & 5142 & \n15.3\tablenotemark{b} & 0.55 & 2.0\\
\enddata
\tablecomments{Column: (1) Designation taken from the Catalog of Isolated Galaxies \citep{Karachentseva1973soobshchspets8_3} or the Sloan Digital Sky Survey \citep{Abazajian2009apjs182_543}, respectively. (2) Abbreviated galaxy designation. (3)-(4) Galaxy coordinates sourced from the NASA/IPAC Extragalactic Database (NED). (5) Adopted distances derived from the Virgo Cluster infall corrected recessional velocities as reported in NED.  (6)  Optical major axis diameter. For the KIG sources, these values come from M04, where they were directly measured from deep imagery; sizes for the rest of the sample are obtained from the diameter measurements listed in NED. (7) Physical major axis diameter derived from Cols. (5) \& (6). (8) Heliocentric optical velocities taken from NED. (9)-(10) Absolute B-band magnitude and B-V color. These data are sourced from M04 for the KIG sample. For the others, photometry was obtained from NED where B \& V-band exists, else transformed using ugriz magnitudes from the NASA Sloan Atlas. Foreground extinction corrections were applied using \citet{Schlafly2011apj737_103}. (11) B-band luminosity in solar units computed from Col. (8), using an absolute solar luminosity M$_{\sun, B} = 5.44$ mag \citep{Willmer2018apjs236_47}.} 
\tablenotetext{a}{Photometry derived through transformation of NSA {\tt Sersic flux} data; see Section~\ref{sect:datastandardization} for details.}
\tablenotetext{b}{S2350 not in the NSA; transformed NSA photometry produced unphysical  B$-$V color for S1323. Photometry derived instead through transformation of SDSS {\tt Cmodel} values using same prescription as for NSA-derived values.}
\end{deluxetable*}

 Basic properties for the IEGs with \HI\ data are given in Table~\ref{tab:galinfo}. The eight IEGs that are the primary focus of this paper are listed in the upper half of the table. Three  of these objects were observed in other \HI\ surveys with two detections: S0926 from early ALFALFA results, \citep{Haynes2011aj142_170} and S1110 \citep{Kreckel2012aj144_16}. S1048 was targeted but not detected by \citet{Kreckel2012aj144_16}. The lower half of the table provides adopted properties for other IEGs having \HI\ data, including observations from Papers~1 \& 2 and from previous surveys that covered five IEGs not observed by us (S1221 \& S1327, \citet{Kreckel2012aj144_16}; S1323, \citet{Most2013aj145_150}, Mrk737, \citet{Salzer2002aj124_191}, and NGC~1211, \citet{Springob2005apjs160_149}).  Unless otherwise noted, these 25 galaxies comprise the ``IEG sample" referenced throughout this paper. These objects span a factor of 400 in luminosity (\n14.2 \textless\ M$_B$ \textless\ \n20.7) and are generally blue (average B\n V $=$ 0.58). Exceptions include KIG684, KIG557, SBS1327+597, KIG792 and KIG824 which have red optical colors (B \n V $\gtrapprox$ 0.9) typical of luminous elliptical galaxies.

\subsection{Comparison Galaxy Sample}\label{sect:compsample}
To place our highly isolated systems in an evolutionary context, the \HI\ properties of the IEGs are compared to those of 86 ETGs drawn from \citet{Huchtmeier1995aap300_675}, \citet{Grossi2009aap498_407} and \citet{Serra2012mnras422_1835, Serra2008aap483_57}  (hereafter, the ``comparison" sample). Data in this comparison sample include both single dish and interferometric observations. Galaxies with known cluster membership (e.g., NGC4406, NGC4472 and NGC4694) were excluded from the comparison sample, as were dwarf companions to large galaxies such as NGC185 and NGC205. Also excluded were galaxies associated with compact groups. The intent of these exclusions is to construct a comparison sample consisting of ``field'' and loose group galaxies. With the exception of the IEG galaxies, the comparison galaxies serve as the lowest density environment sample with published \HI\ data. A published \HI\ observation for another galaxy, NGC7465, was removed because the beam included gas from other group members. All other \HI-detected ETGs from these four references were included. Duplication was eliminated by using data from the source having the most recent publication date.  Over 90\% of these galaxies are catalogued group members.  The remaining $\sim$10\% have no apparent catalogued group/cluster affiliation and are considered to be ``field'' galaxies. A search of their environments using NED\footnote{The NASA/IPAC Extragalactic Database (NED) is operated by the Jet Propulsion Laboratory, California Institute of Technology, under contract with the National Aeronautics and Space Administration.} reveals that most of the field galaxies have nearest-neighbors within 0.5~Mpc. The IEGs differ dramatically from the comparison galaxies with respect to degree of isolation. Only one field galaxy in the comparison sample, NGC6798 \citep{Serra2012mnras422_1835}, appears to have a low-density environment comparable to the IEGs. Table~\ref{tab:compinfo} lists the optical and \HI\ properties for the comparison sample (see Appendix~\ref{sect:compdata}). 


\section{Observations and Data Reduction}\label{sect:obsdatareduction}

\subsection{Data Standardization}\label{sect:datastandardization}
Properties including optical photometry and distances for both the IEG and comparison samples were extracted from several different published surveys and required  standardization. Distances were Virgo Infall corrected assuming the following adopted parameters: H$_0=73.0$, $\Omega_{\rm{matter}}=0.27$, $\Omega_{\rm{vacuum}}=0.73$. The $B-$band photometry and $B\n V$ colors were adopted in the following order of availability:  our own ground-based data \citep{Marcum2004aj127_3213}, $B_{T}$ (RC3), $m_{B}$ (RC3), or photometry derived from ugriz fluxes of 2-D Sersic model fits (specifically, the {\tt NMGY photometry}) taken from the NASA-Sloan Atlas (http://www.nsatlas.org/, hereafter ``NSA'') and transformed to the Johnson filter system using the prescription of \citet{Cook2014mnras445_890}. Optical photometry is extinction-corrected using \citet{Schlafly2011apj737_103}. Tables~\ref{tab:galinfo} and \ref{tab:compinfo} indicate photometry transformed from the NSA.

Values for \HI\ fluxes for the comparison sample, when directly provided by the source, were used to derive \HI\ masses. If only \HI\ masses were given in a paper, then associated \HI\ fluxes were inferred using distances adopted by the reference. 

\subsection{\HI\ Observations}\label{sect:hiobs}
\HI\ observations were made using the 100m Robert Byrd Green Bank Telescope from October through December 2017. The GBT provides a 9\arcmin\ beam which is sufficiently sized to detect emission from neutral gas reservoirs within or surrounding the IEG sample. Using the L-band receiver and {\tt VEGAS} backend, position-switched single pointing \HI\ observations were made over a bandwidth of 16.875 MHz.  Calibrators were chosen from the GBT calibrator catalog during the beginning of each observing run and were used to point and focus the telescope.

Standard data reduction techniques were used to calibrate the data using GBTIDL\footnote{Developed by NRAO; for documentation see \url{http://gbtidl.nrao.edu/}.}.  The data (with an original channel width of 0.22 \kms) were Hanning smoothed and the procedure \textsc{getps} was used to calibrate the antenna temperature of the data. RFI spikes were identified and manually removed by interpolating over neighboring channels.  Boxcar smoothing over three channels was performed to increase the signal-to-noise ratio of the line profiles.  Polynomials of 3rd-7th order were then fit to the baselines in order to remove environmental and instrumental effects.  Finally, the data were smoothed using a boxcar function over five channels resulting in a final channel width of \s6.9 \kms.

The GBTIDL procedure \textsc{stats} was used to derive an estimate of the \HI\ mass and root-mean-square (rms) uncertainty for each galaxy.  The noise rms was calculated using emission-free baseline-subtracted channels.  The \textsc{stats} procedure is used to measure the area under the curve of any emission in the line profiles, a value equal to the sum of flux (S) over N channels times the channel width. This integrated flux density is used in the following equation to calculate an \HI\ mass: 

\begin{equation}\label{eqn:mass}
M_{\rm HI}(M_{\sun})=2.356 \times 10^5\ D^{2}\sum_{i}S_{i}\Delta V
\end{equation}
where D is the distance of the galaxy in Mpc, S$_{i}$ is the flux in the $i^{th}$ channel in Jy, and $\Delta V$ is the channel width in \kms.

\begin{deluxetable*}{lccccccrc}
\tablecaption{GBT Observations\label{tab:gbttable}}
\tabletypesize{\footnotesize}
\tablecolumns{9}
\tablewidth{0pt}
\tablehead{
\colhead{Name} & \colhead{T$_{source}$}  & \colhead{Physical Beam} & \colhead{rms} & \colhead{S$_{\rm{HI}}$} & \colhead{$\Delta$ V} & \colhead{M$_{\rm{HI}}$} &  \colhead{log$_{10}$(M$_{\rm{HI}}$/L$_{\rm{B}}$)} & \colhead{\HI\ Source}\\
\colhead{} & \colhead{(hr)} & \colhead{(kpc)} & \colhead{(mJy)} & \colhead{(Jy \kms)} & \colhead{(\kms)} & \colhead{(10$^8$ M$_{\sun}$)} & \colhead{(M$_{\sun}$/L$_{\sun}$)} & \colhead{}\\
\colhead{(1)} & \colhead{(2)} & \colhead{(3)} & \colhead{(4)} & \colhead{(5)} & \colhead{(6)} & \colhead{(7)} & \colhead{(8)} & \colhead{(9)}}

\startdata
KIG164 & 2.0 & 333 & 0.64 & 1.13 $\pm$ 0.03 & 340 & 43 $\pm$ 1 & \n0.74 & 1 \\
SDSSJ092657.99+100301.4 & 3.9 & 191 & 0.49 & 1.83 $\pm$ 0.02 & 200 & 23.0 $\pm$ 0.2 & \n0.02 & 1 \\
SDSSJ094410.80+001047.3 & 3.5 & 120 & 0.51 & 0.35 $\pm$ 0.02 & 145 & 1.7 $\pm$ 0.1 & \n0.53 & 1 \\
SDSSJ104807.06+430525.5 & 4.5 & 152 & 0.49 & 0.62 $\pm$ 0.02 & 290 & 4.9 $\pm$ 0.2 & +0.20 & 1 \\
SDSSJ111029.62+134558.1 & 4.5 & 159 & 0.38 & 0.39 $\pm$ 0.01 & 160 & 3.4 $\pm$ 0.1 & \n0.34 & 1 \\
KIG684 & 5.6 & 205 & 0.41 & $\le$0.05 & \nodata & $\le$0.8 & $\le$\n2.08 & 1 \\
SDSSJ155325.18+520416.5 & 2.0 & 125 & 0.60 & 0.66 $\pm$ 0.02 & 200 & 3.5 $\pm$ 0.1 & \n0.40 & 1 \\
SDSSJ212753.52\n070113.9 & 2.0 & 129 & 0.65 & 1.90 $\pm$ 0.02 & 190 & 10.9 $\pm$ 0.1 & \n0.28 & 1 \\
\cutinhead{IEGs HI Data Presented in Other Papers}
SDSSJ024248.69\n082356.6 & 9.0 & 152 & 0.49 & $\le$0.06 & \nodata & $\le$0.5 & $\le$\n1.46 & 3\\
NGC1211 & \nodata & 113 & \nodata & 14.98 & \nodata & 65.4 & \n0.43 & 7\\
VIIIZw40 & 6.0 & 132 & 0.86 & $\le$0.11 & \nodata & $\le$0.7 & $\le$\n1.29 & 2\\
SDSSJ102145.89+383249.8 & 7.6 & 158 & 0.62 & 0.60 $\pm$ 0.03 & 260 & 5.1 $\pm$ 0.2 & \n0.32 & 2\\
Mrk150 & 7.5 & 141 & 0.65 & 1.34 $\pm$ 0.03 & 280 & 9.2 $\pm$ 0.2 & \n0.54 & 3\\
SDSSJ113237.43+472658.7 & 2.5 & 65 & 0.99 & 1.24 $\pm$ 0.03 & 170 & 1.8 $\pm$ 0.0 & +0.39 & 3\\
Mrk737 & \nodata & 128 & \nodata & 1.49 & \nodata & 8.4 & \n0.15 & 6\\
SDSSJ122123.13+393659.5 & \nodata & 170 & \nodata & 0.35 & \nodata & 3.5 & +0.02 & 4\\
KIG557 & 8.6 & 515 & 0.78 & $\le$0.10 & \nodata & $\le$9.2 & $\le$\n1.47 & 2\\
SDSSJ132337.69+291717.1 & \nodata & 155 & \nodata & 0.17 & \nodata & 1.4 & \n0.28 & 5\\
SBS1327+597 & \nodata & 189 & \nodata & 2.07 & \nodata & 25.4 & +0.02 & 4\\
SDSSJ143538.00+435937.1 & 11.0 & 202 & 0.51 & $\le$0.07 & \nodata & $\le$0.9 & $\le$\n1.00 & 2\\
KIG792 & 6.5 & 339 & 0.62 & $\le$0.08 & \nodata & $\le$3.2 & $\le$\n1.82 & 2\\
KIG824 & 7.5 & 204 & 0.51 & 0.32 $\pm$ 0.02 & 330 & 4.6 $\pm$ 0.3 & \n1.33 & 3\\
KIG870 & 5.9 & 234 & 0.53 & 2.91 $\pm$ 0.03 & 475 & 54.8 $\pm$ 0.6 & \n0.47 & 2\\
SDSSJ234042.69\n092336.6 & 6.8 & 184 & 0.55 & 0.28 $\pm$ 0.02 & 160 & 3.3 $\pm$ 0.2 & \n0.95 & 3\\
SDSSJ235021.45+141342.6 & 5.2 & 187 & 0.59 & 1.30 $\pm$ 0.02 & 220 & 15.7 $\pm$ 0.3 & +0.90 & 3\\
\enddata
\tablecomments{Column: (1) Galaxy designation. (2) Total on-source time. (3) The projected 9\arcmin\ FWHM values of the GBT beam at the distances provided in Table~\ref{tab:galinfo}. (4) Measured rms noise per channel. (5) \HI\ flux density with 1$\sigma$ uncertainties. Upper limits (5$\sigma$) are given for nondetections. (6) Full width at zero intensity of the emission profile, defining the velocity range over which the GBTIDL program \textsc{stats} was used to obtain an \HI\ integrated flux density. Upper limit \HI\ masses use $\Delta$V$=$100 \kms. (7) Derived \HI\ mass with 1$\sigma$ uncertainties. (8) Log of the \HI\ mass-to-blue luminosity ratio in solar units. (9) Source of \HI\ data: 1-- this paper; 2--\citet{Ashley2018aj155_15}; 3--\citet{Ashley2017aj153_158}; 4--\citet{Kreckel2012aj144_16}; 5--\citet{Most2013aj145_150}; 6--\citet{Salzer2002aj124_191}; 7--\citet{Springob2005apjs160_149}}.
\end{deluxetable*}

\begin{figure*}[h!] 
\epsscale{.49}
\plotone{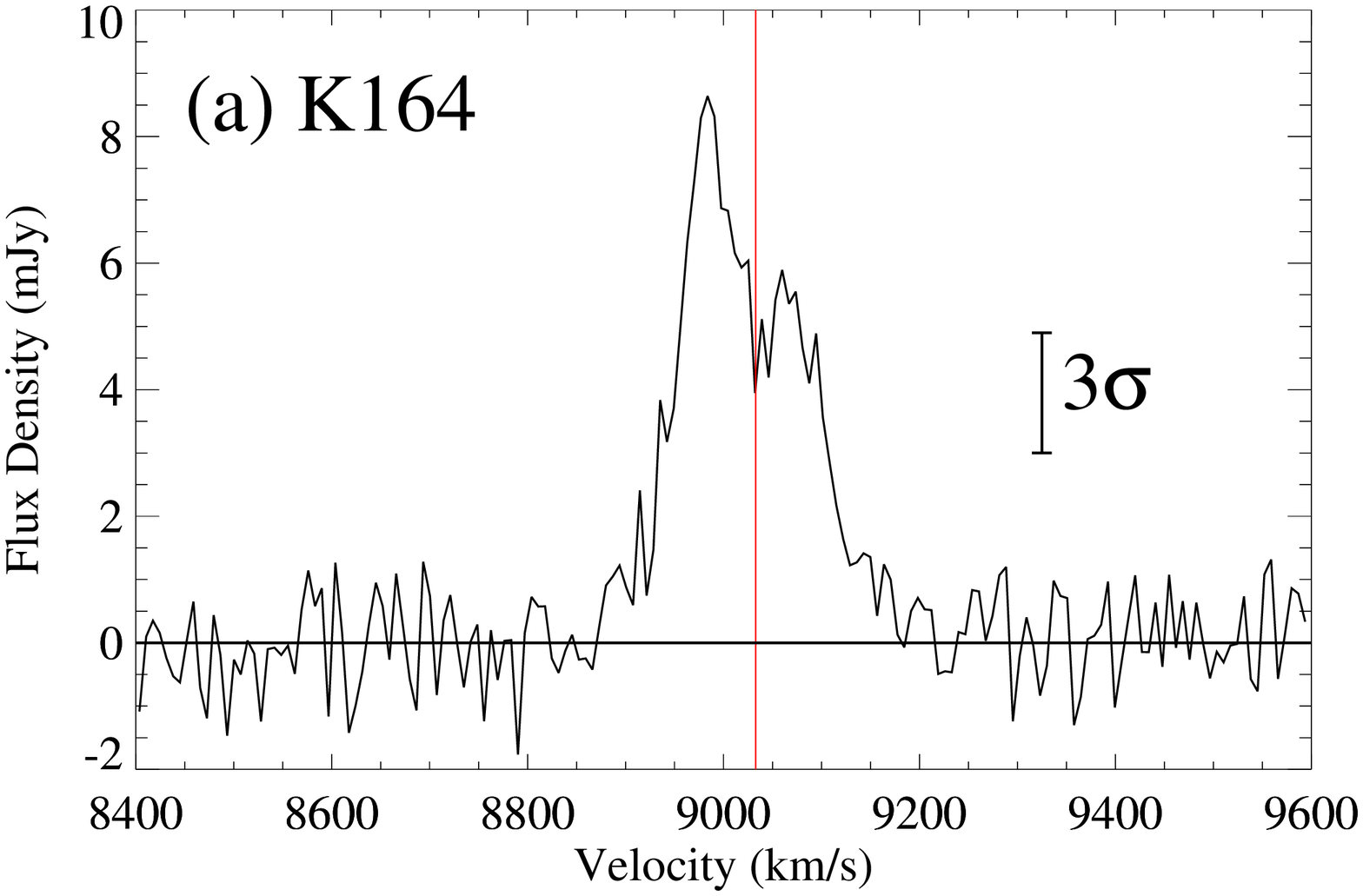}
\plotone{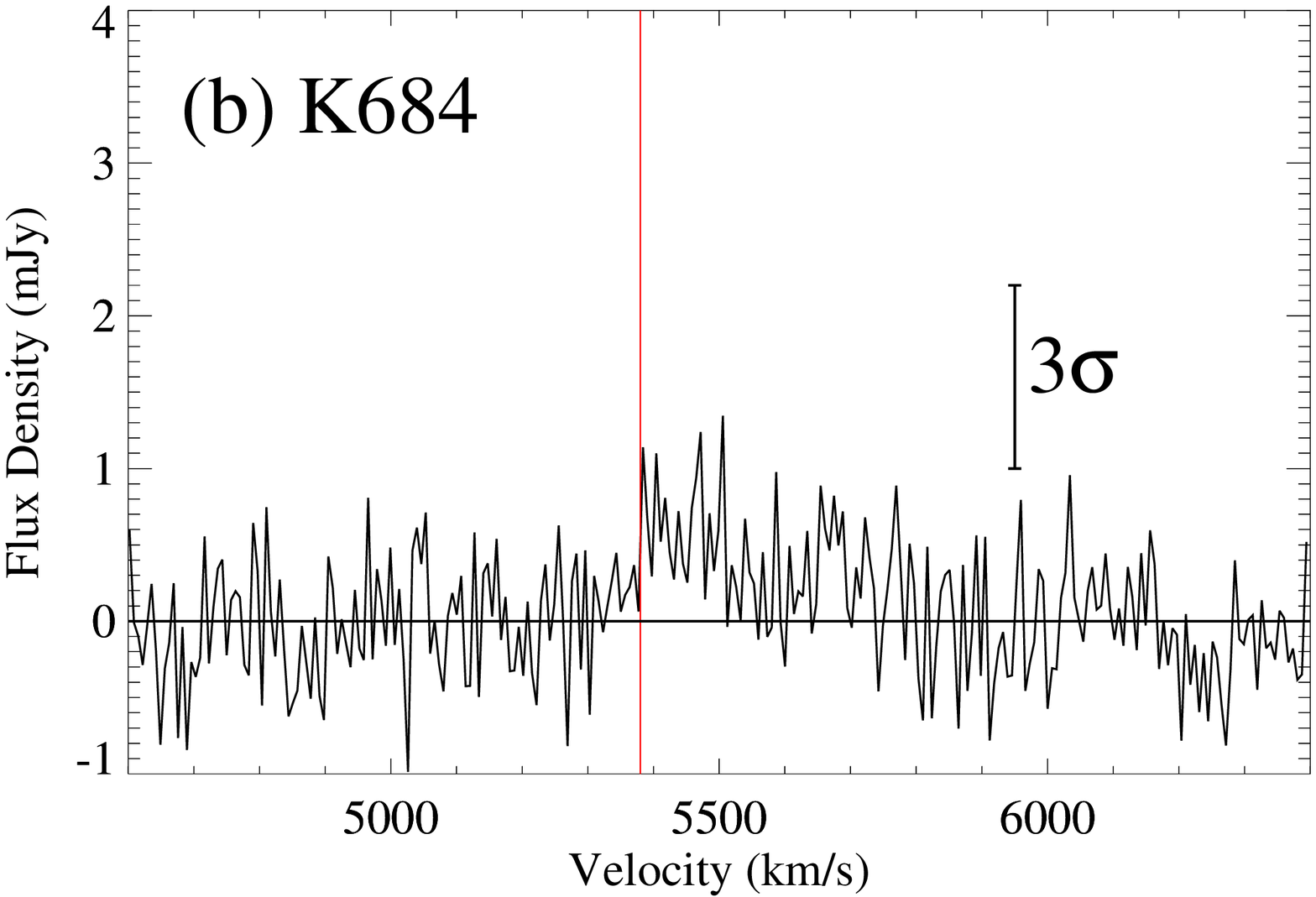}\\
\plotone{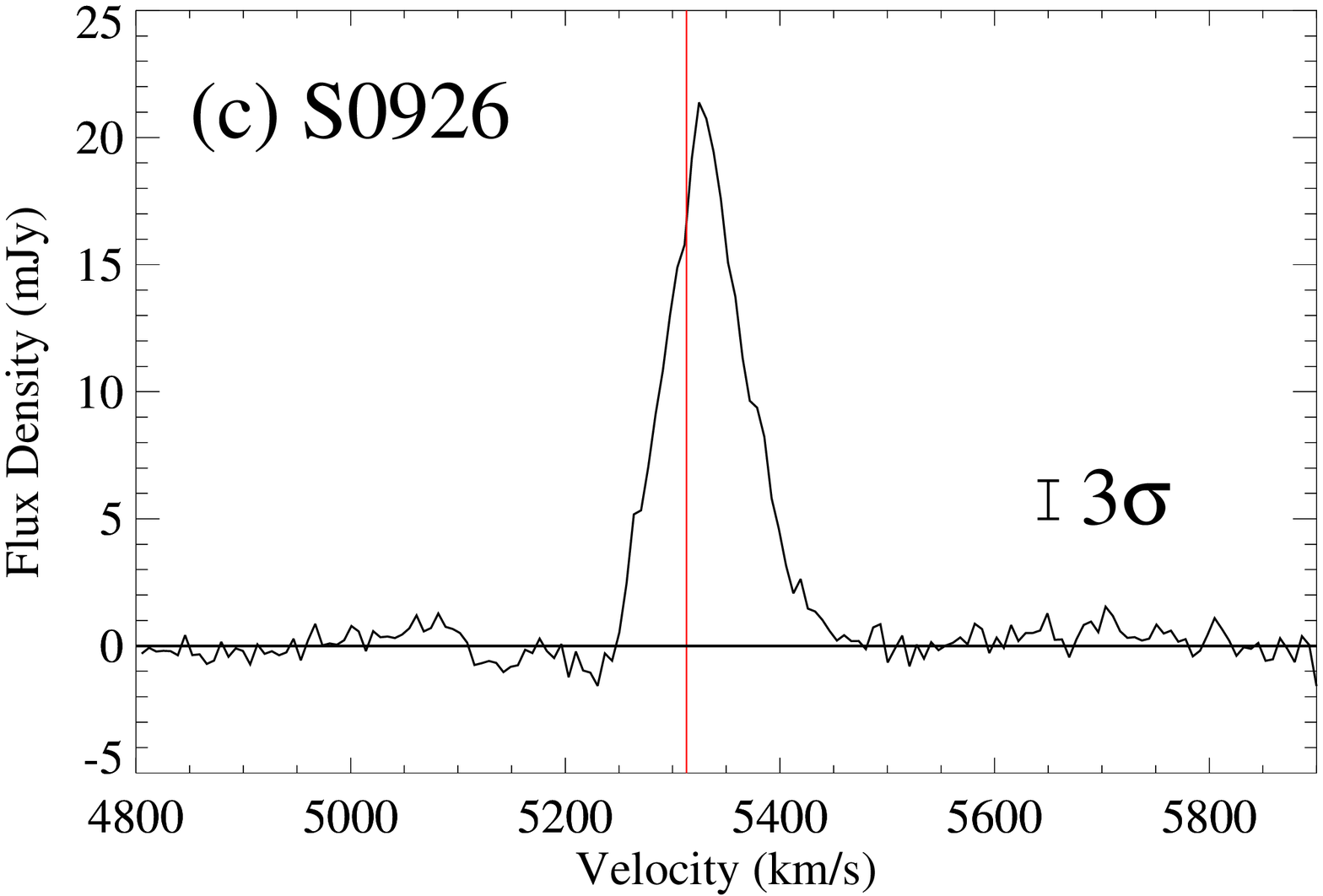}
\plotone{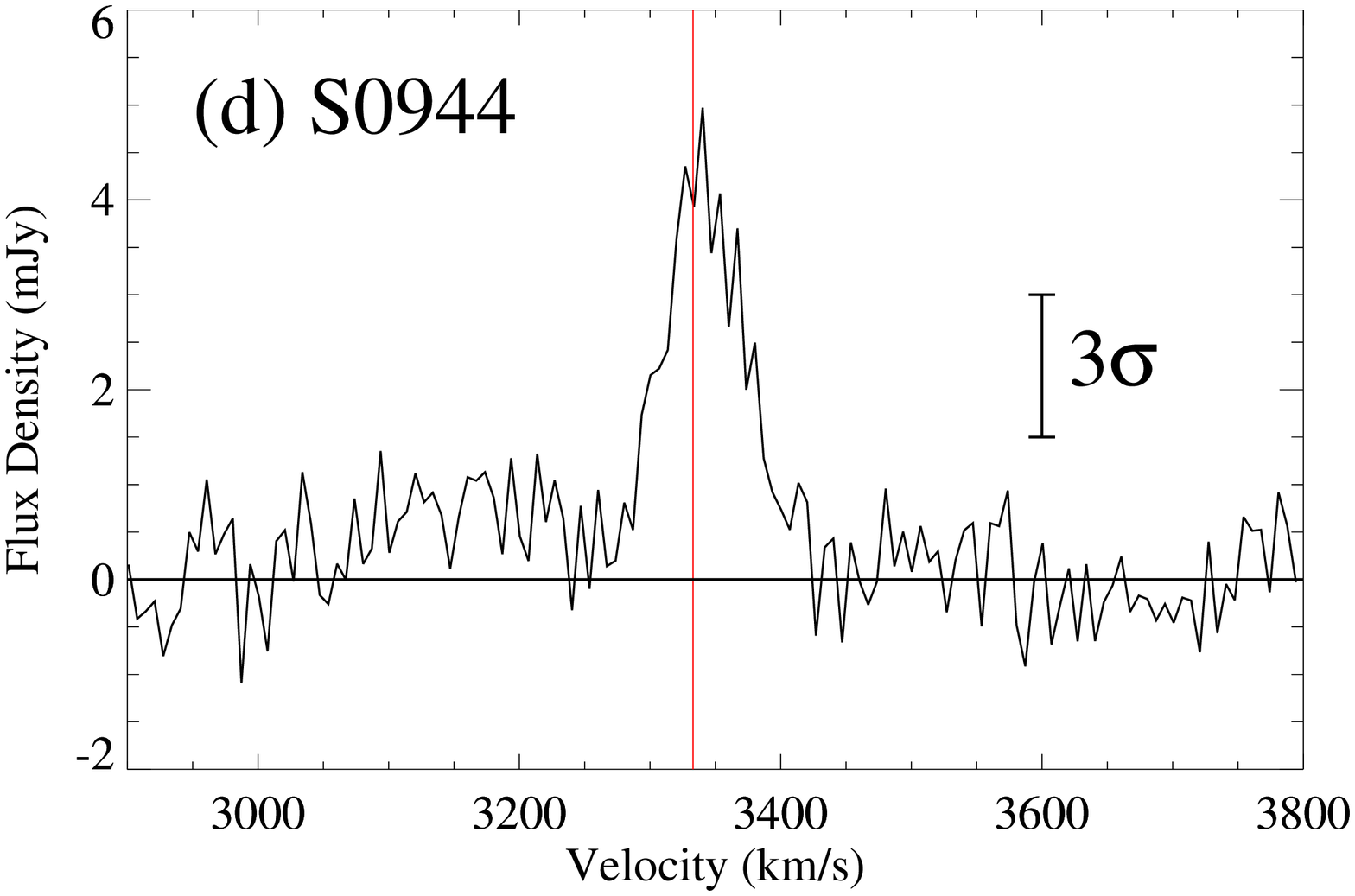}\\
\plotone{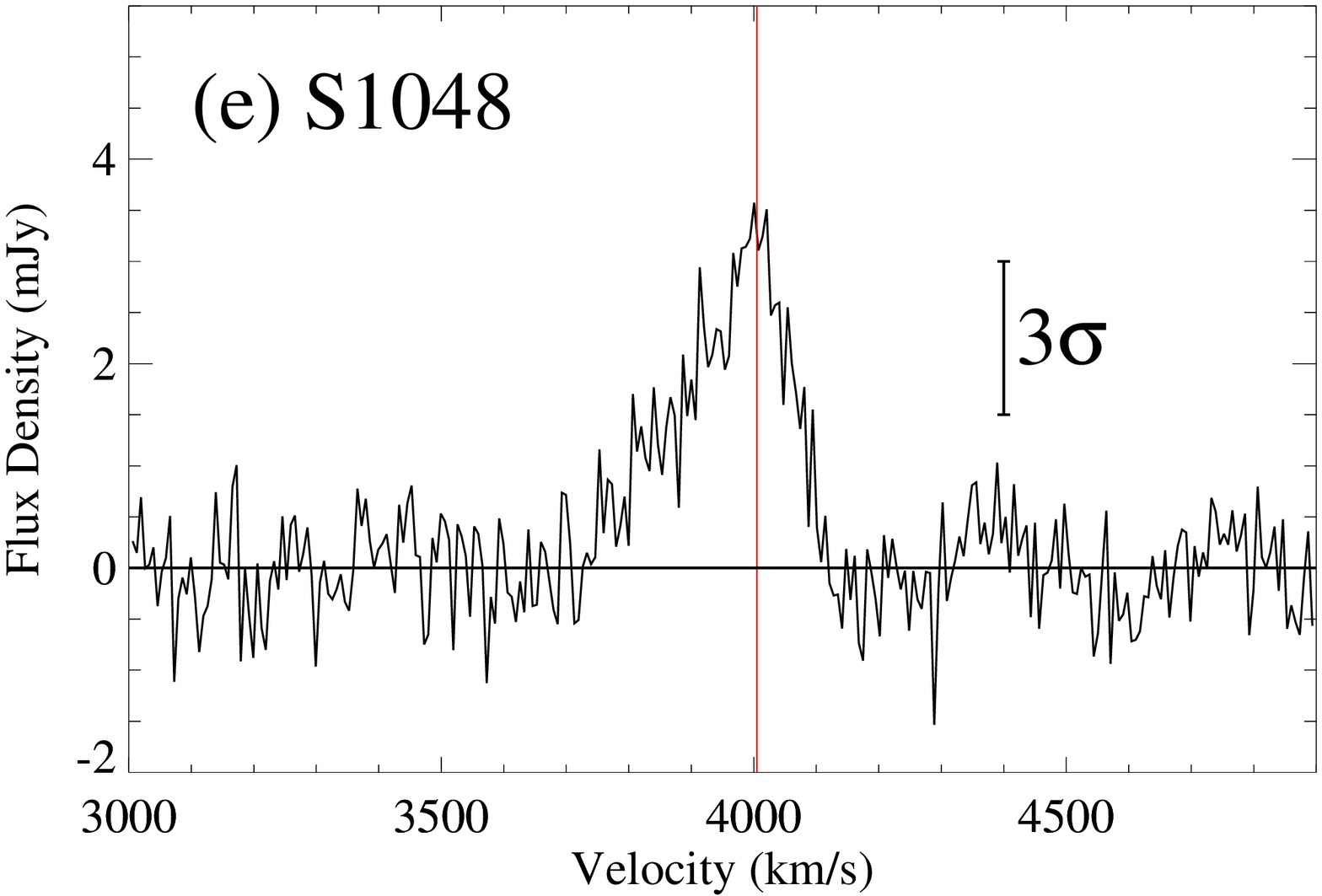}
\plotone{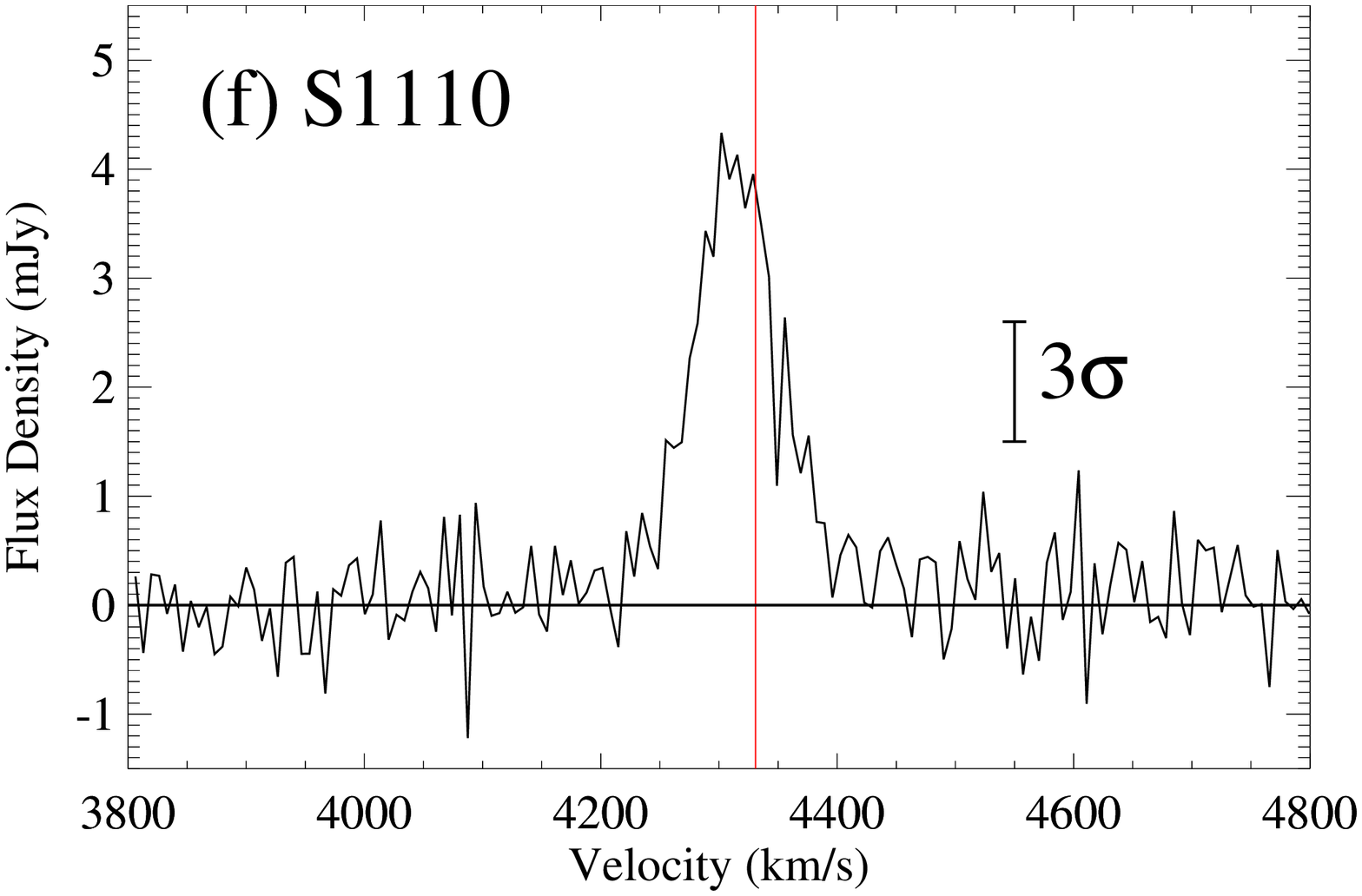}\\
\plotone{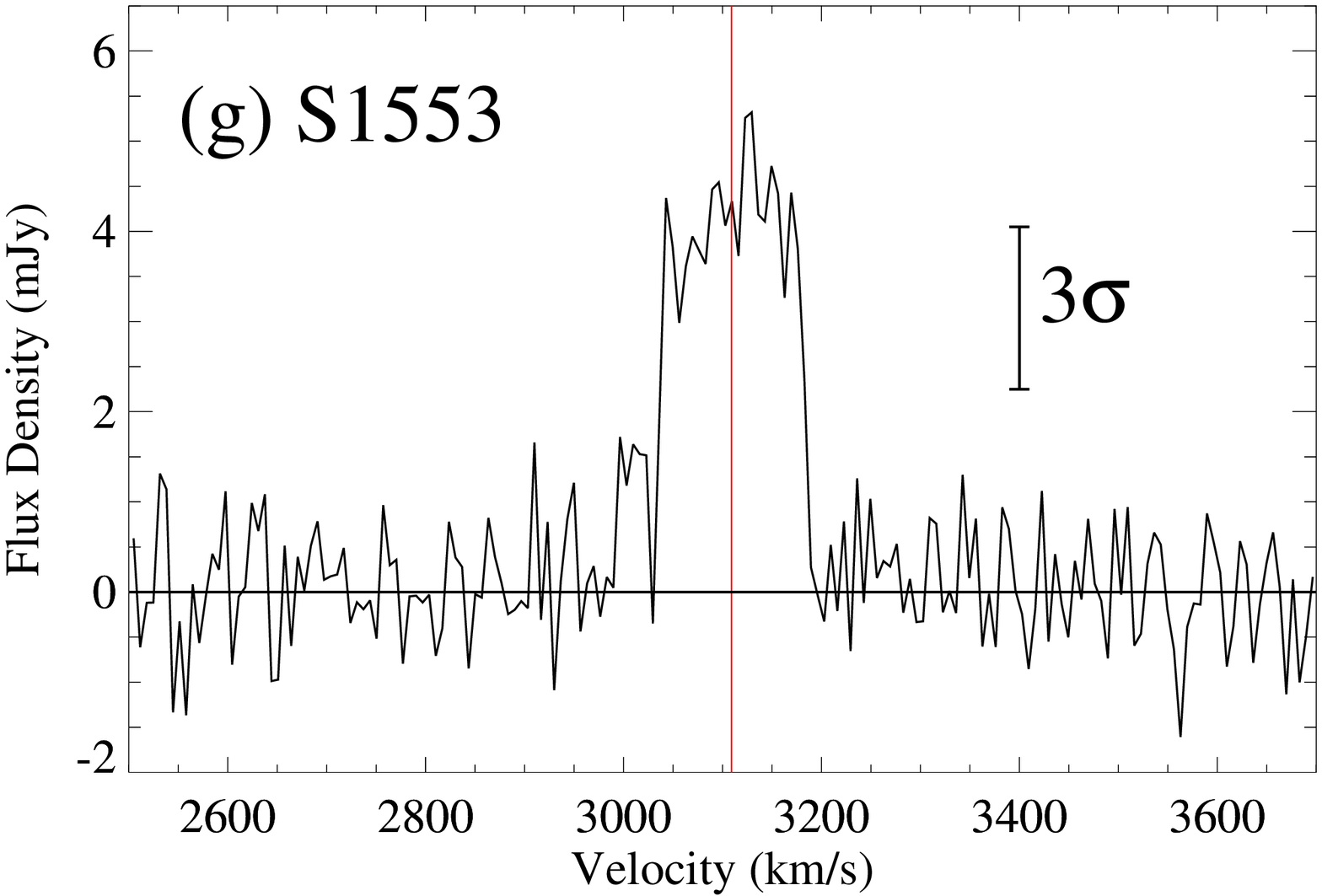}
\plotone{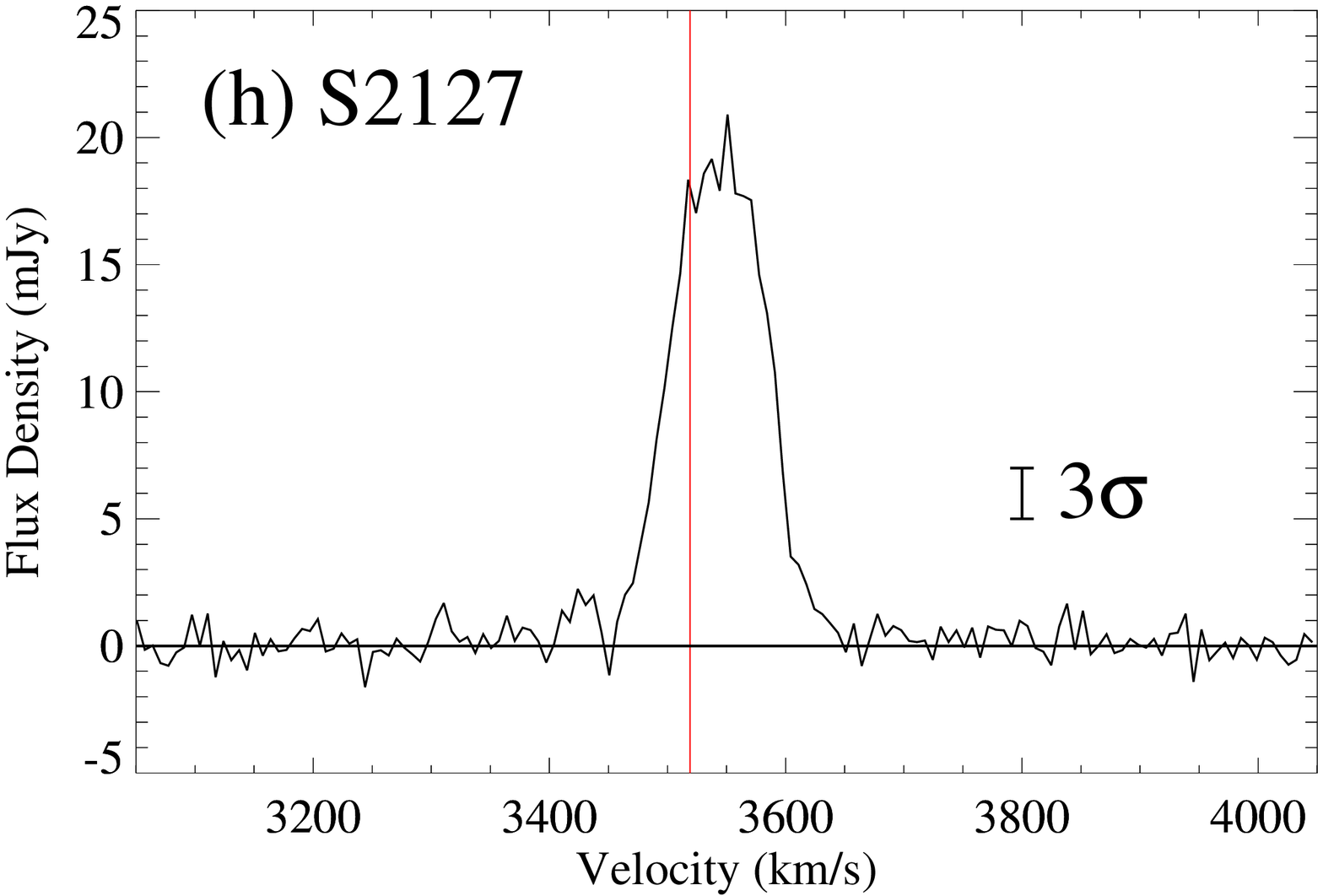}
\caption{\HI\ line profiles for the 8 IEGs reported here. Source names are given in the upper left corner, shortened for brevity. The red vertical lines indicate the optically defined systemic recessional velocity for each galaxy, sourced from NED. Error bar lengths are 3 times the rms per channel.  These profiles have been Hanning and boxcar-smoothed resulting in a 6.9 \kms\ channel width.
}
\label{fig:lineprof}
\end{figure*}

Table~\ref{tab:gbttable} provides information associated with the \HI\ observations, measured values, and and derived values for each IEG (the upper half contains the most recent observations, the lower half contains observations from the literature\footnote{Note that some \HI\  and photometric values for galaxies from Papers~1 \& 2 may be slightly different from their previously-published values due to differences in adopted distances.}).  Upper limits for \HI\ masses are calculated by setting the \HI\ flux density to five times the rms per channel ($\sigma_{rms}$), assuming that the emission is spread over 100 \kms\ \citep[a number that is easy to adjust and a reasonable velocity width for ETGs:][]{Huchtmeier1995aap300_675, Grossi2009aap498_407}: 

\begin{equation}\label{eqn:upperlim}
M_{\rm{HI}}\rm(upper)=2.356 \times 10^5\ D^{2}(5\sigma_{rms} \times \frac{100}{\sqrt{N}}).
\end{equation}
where N is the number of channels spanning 100 \kms.
Figure~\ref{fig:lineprof} presents the \HI\ line profiles of the IEG sample.

\section{Results and Discussion}\label{sect:resultanddiscussion}

\subsection{Individual Galaxies}\label{sect:ind_galaxies}

Four of the \HI\ spectra shown in Figure~\ref{fig:lineprof} show single-peaked Gaussian-like shaped profiles. Two galaxies exhibit asymmetric emission around the line centroid and one shows a steep-sided rectangular shape, similar to the double-horned profile typical of rotating disk systems but lacking the central deficit of emission which produces the ``horns." Although a unique determination of the \HI\ gas distribution is not possible using just the single dish spectrum, the shape of the profile can be used to obtain information about the global gas content  and gas kinematics.  

A number of plausible 3D \HI\ distributions may produce the observed spectral shapes. Gaussian-like profiles may indicate that the \HI\ has little rotational support or is contained in a low-inclination disk with little line-of-sight velocity spread. Lower luminosity systems may possess dwarf-like disks with near solid body rotation and a high central gas density producing a Gaussian-like profile. \citet{Serra2012mnras422_1835} found that roughly two-thirds of their ETG sample detected in \HI\ (64\%) have rotating \HI\ discs or rings. Since their sample consisted of luminous ETGs, we would expect that our more luminous systems also contain a significant fraction of rotating disks. For most of our IEGs, the emission velocity widths displayed in Figure~\ref{fig:lineprof} are sufficiently large which could be the result of substantial rotational support. Profiles may also be broadened and exhibit asymmetries due to tidally (merger) induced dynamic perturbations, producing both infalling and outflowing clouds. Both gas-rich KIGs in our sample show optical-band distortions in their 2D light profiles \citep{Marcum2004aj127_3213}, strongly suggesting that these systems have accreted gas from a past merger. Finally, while these systems are all isolated from nearby massive companions, faint \HI-rich dwarf companions or circumgalactic material lying within the 9\arcmin\ GBT beam could contribute to profile asymmetry.  

Below we briefly discuss each system, more profile analysis will be presented in Paper~IV.   

\begin{itemize}
\item[] {\bf KIG164} (Figure~\ref{fig:lineprof}a) shows strong asymmetric emission with two peaks having amplitudes reaching 14$\sigma$ and 9$\sigma$, respectively. The optical systemic velocity falls directly within the dip in amplitude between the two peaks. The profile suggests that more gas is blueshifted with respect to the systemic optical velocity than is redshifted. KIG164 possesses the second largest \HI\ mass in the larger 25 IEG sample of this work with M$_{\rm HI}$=$4.7\times10^9$~M$_\sun$, only surpassed by KIG870 which has an \HI\ mass of $5.8\times10^9$~M$_\sun$.  This \HI\ mass is also large relative to that of the ETG samples of \citet{Grossi2009aap498_407} and \citet{Serra2012mnras422_1835}. Only two galaxies in those studies have larger \HI\ masses: UGC4599 and NGC7465 with \HI\ masses of $7.6 \times10^9$~M$_\sun$ and $9.5\times10^9$~M$_\sun$, respectively. Galaxies in the \citet{Huchtmeier1995aap300_675} sample have larger \HI\ masses than that of Grossi and Serra, with six galaxies containing \HI\ masses greater than that of KIG164 (M$_{\rm{HI}}$ of $5.3-18.7\times10^9$~M$_\sun$). 

Although the \HI\ profile displays two peaks, the profile's asymmetry, and narrow and shallow dip together do not match the expected characteristics of a rotating disk with a double-horn profile.  Using deep optical imagery \citet{Marcum2004aj127_3213} found low-level features indicative of an advanced merger, including a morphologically asymmetric nucleus and a potential stellar bridge between KIG164 and a possible dwarf companion. KIG164's \HI\ line profile could indicate that the gaseous component of the galaxy is still in the process of kinematically relaxing into a rotating disk following a recent accretion event or an interaction with the dwarf. KIG164 and KIG870 (discussed in Paper 2) are both gas-rich, blue (B\n V $= 0.47$, 0.48, respectively) and luminous (M$_{\rm{B}}$ = \n20.5, \n20.1, respectively), far removed from the red sequence of luminous ETGs in the galaxy color-magnitude diagram.  Both galaxies exhibit subtle photometric evidence \citep{Marcum2004aj127_3213} for an interaction and have high-mass star formation as measured by their very blue B\n V colors.  

\item[] \textbf{KIG684} (Figure~\ref{fig:lineprof}b) has no reliably detected emission at the sensitivity of the data. The spectrum does show potential low-level emission with a plateau centered on \s5425 \kms, close to the galaxy's optically-defined systemic velocity. Due to the low significance of this feature  (\s1.4$\sigma$), we regard this galaxy as being a non-detection in \HI\ emission. 

\item[] \textbf{SDSSJ0926} (Figure~\ref{fig:lineprof}c) has very strong emission with a single peak detected with an SNR of 44$\sigma$.  \citet{Haynes2011aj142_170} observed this system as part of the Arecibo ALFALFA survey and finds a similarly-shaped Gaussian-like profile with an \HI\ centroid at 5334 \kms, and an integrated flux density \s6.5\% larger than ours, consistent with the GBT data. The optical systemic velocity 5313 \kms\ ($\pm$2 \kms) is slightly lower than the velocity of the \HI\ peak at 5325 \kms.  S0926 contains a large \HI\ mass of $2.2\times10^9$ M$_\sun$. \citet{Fuse2012aj144_57} indicate that S0926 may have stellar tidal arms.  The \HI\ line profile does not show any obvious signs of this disturbance.  The gaseous tidal arms might not show as prominent features in an \HI\ line profile if they are rotating with the disk. Alternatively, the gaseous tidal arms may have changed gas phase \citep{Mihos2001apj550_94} or kinematically decoupled from the stellar component \citep{Duc1997aap553_537} leaving a residual stellar tidal feature. 

\item[] \textbf{SDSSJ0944} (Figure~\ref{fig:lineprof}d) has a Gaussian-like \HI\ profile shape similar to that of S0926 with a peak emission detected with an SNR of 8$\sigma$.  The optical systemic velocity and velocity corresponding to the \HI\ line profile centroid are in good agreement. No optically interesting features were noted in \citet{Fuse2012aj144_57}.

\item[] \textbf{SDSSJ1048} (Figure~\ref{fig:lineprof}e) exhibits an asymmetric Gaussian-like profile with the peak emission having an SNR of \s7$\sigma$ and matching well with the optical systemic velocity.  S1048's \HI\ line profile is similar in shape to those of S0926 and S0944, however, the shape of the profile is noticeably asymmetric about the optical systemic velocity with a bulk of the emission skewed towards blue-shifted velocities. \citet{Fuse2012aj144_57} do not identify any optical disturbances in this system that could be associated with the asymmetry in the \HI\ line profile. 
SDSSJ1048, also designated VGS16, was observed but \textit{not} detected by \citet{Kreckel2012aj144_16} as part of their Void Galaxy Survey using the Westerbork Synthesis Radio Telescope (WSRT). Their 3$\sigma$ \HI\ mass upper limit is $1.16\times10{^8}$ M$_{\sun}$, computed for an assumed profile width of 100 \kms.  This upper limit is consistent with the measured mass provided in Table \ref{tab:gbttable} after taking into account the broad detected profile width (\s290 \kms).

\item[] \textbf{SDSSJ1110} (Figure~\ref{fig:lineprof}f) has an  emission profile with a Gaussian-like shape. The profile peak has a SNR of  \s11$\sigma$ and lies at 4310 \kms\ compared to the optical systemic velocity of 4331 \kms ($\pm$25 \kms).  This system was observed by \citet{Kreckel2012aj144_16} designated VGS19. Using WSRT, they report an \HI\ systemic velocity centroid at 4318 \kms\ and an \HI\ mass of $3.16\pm0.76 \times10^{8}$  M$_{\sun}$, consistent with the measurements presented here.

\item[] \textbf{SDSSJ1553} (Figure~\ref{fig:lineprof}g) has an \HI\ line profile with steep edges and a noisy plateau, giving the profile a rectangular shape.  The plateau is detected at an SNR of \s7$\sigma$ with the optical systemic velocity falling near the center of the plateau.  The rectangular profile could be a shallow double-horned profile that is obscured by the noise. A small second plateau is centered near 3010 \kms.  This second feature was included in the \HI\ mass measurement in Table~\ref{tab:gbttable} although it is observed with an SNR of only \s2.3$\sigma$.  Excluding the emission contribution from this feature, the \HI\ mass decreases from 2.7 to 2.5$\times10^8$ M$_\sun$.

\item[] \textbf{SDSSJ2127} (Figure~\ref{fig:lineprof}h) has a Gaussian-like shaped \HI\ line profile similar to those of S0944, S0926, and S1110 with a peak emission having an SNR of 28$\sigma$.  The profile peak  falls at \s3545 \kms, noticeably higher than the optical systemic velocity of 3519 \kms\ ($\pm$5 \kms).
\end{itemize}

\subsection{Gas Mass-to-Blue Light Ratios}\label{sect:gas2mass}

\begin{figure*}[h!]
\epsscale{0.8}
\plotone{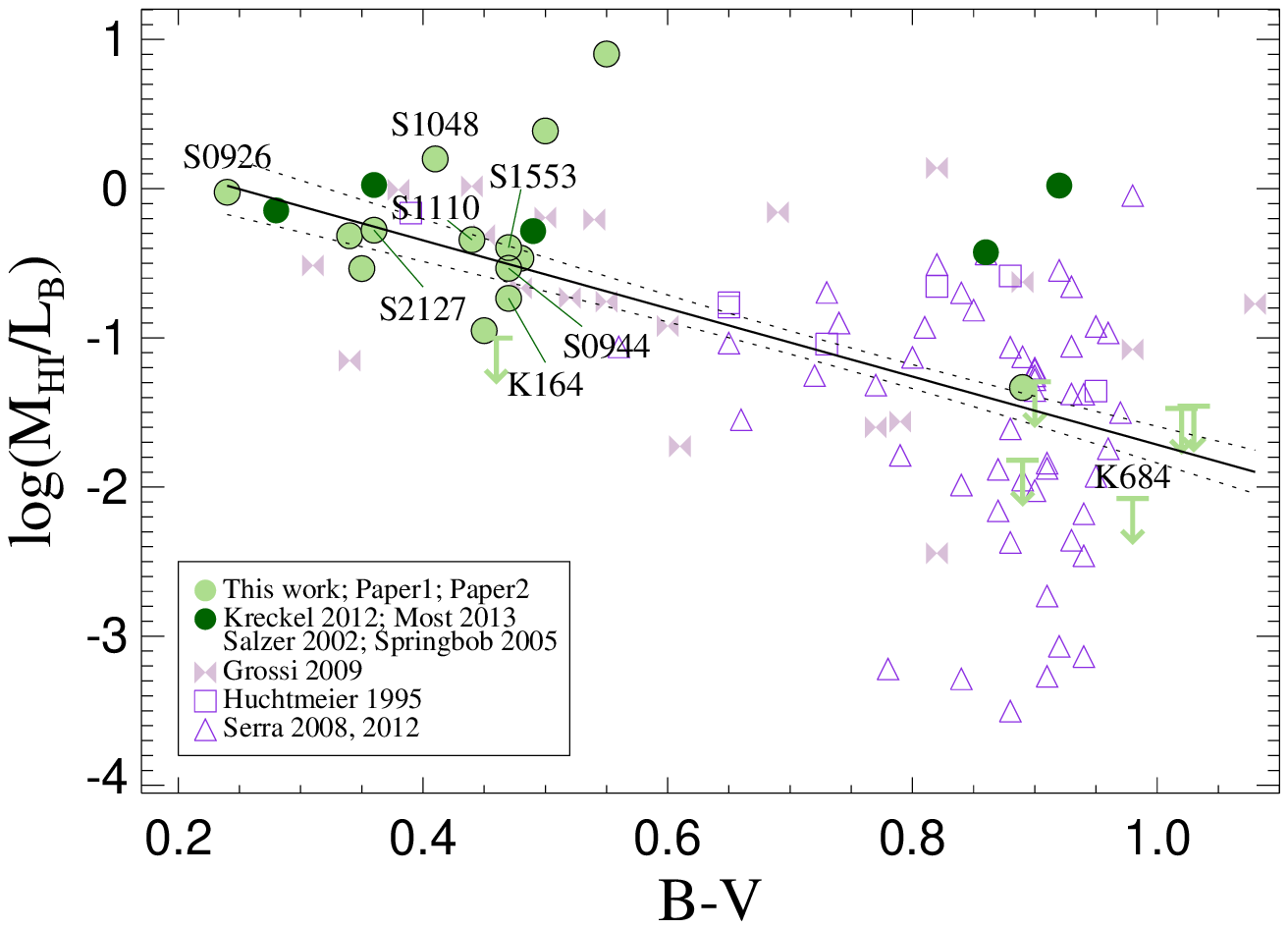}
\caption{ B\n V vs. log(M$_{\rm{HI}}$/L$_{\rm{B}}$) for the entire IEG sample contrasted with the comparison sample. Data presented here and in \citet{Ashley2017aj153_158, Ashley2018aj155_15} are shown in filled light green circles. Data for IEGs Mrk737, SDSSJ122123.13+393659.5, SBS1327+597, SDSSJ132337.69+291717.1 and NGC1211 obtained from, respectively \citet{Salzer2002aj124_191},  \citet{Kreckel2012aj144_16} (2), \citet{Most2013aj145_150} and \citet{Springob2005apjs160_149} are displayed in dark green. Data points centered directly under, above or linked to abbreviated galaxy names indicate the 8 new objects presented in this paper. Green arrows represent $5\sigma$ upper mass limits for non-detections of IEGs. The comparison ETG sample  is shown in purple with the symbol indicating the published source. The linear relationship (black solid line) includes both the detected IEGs and comparison galaxies; dashed lines show the $\pm1\sigma$  envelop.}
\label{fig:logMHILB_BV}
\end{figure*}

Papers 1 \& 2 discuss the relationship between  \HI\ mass-to-blue luminosity ratio (M$_{\rm{HI}}$/L$_{\rm{B}}$) and optical (B\n V) color in ETGs within low-density environments. Figure~\ref{fig:logMHILB_BV} displays this relationship with our IEG sample shown as green circles. The eight IEGs from this paper are indicated by their abbreviated name. Non-detections are shown as 5$\sigma$ upper limits. The five IEG galaxies having \HI\ measurements found in the literature \citep{Salzer2002aj124_191, Kreckel2012aj144_16, Most2013aj145_150, Springob2005apjs160_149} are indicated by dark green circles. Galaxies from the comparison sample are displayed in purple. As noted in Papers 1 \& 2, Figure~\ref{fig:logMHILB_BV} shows that bluer galaxies are generally more gas-rich (higher M$_{\rm{HI}}$/L$_{\rm{B}}$) than their red counterparts. We found this relationship to be sufficiently robust to provide predictions of \HI\ flux as the basis of exposure time estimates for GBT observations of the IEGs, most of which had not previously been observed in \HI. 
The solid line in Figure~\ref{fig:logMHILB_BV}, a linear fit to all of the data excluding upper limits, is given as: 
\begin{equation}\label{eqn:fit2}
log(M_{\rm{HI}}/L_{\rm B}) = -2.3(B-V) + 0.6.
\end{equation}

Figure~\ref{fig:logMHILB_BV} illustrates the large range in B\n V color for the 25 IEGs: $\sim$0.25 to $\sim$1.0. The IEGs fill in the otherwise sparsely-populated blue end of the \HI\ mass/color relationship. The M$_{\rm{HI}}$/L$_{\rm{B}}$ for IEGs with detected \HI\ range from $\sim10^{-4}$ to 8.  Of the comparison galaxies, the Grossi sample has characteristics most similar to those of the IEGs; the galaxies in that study are bluer and more dwarf-like in size than other ETGs. The majority of null detections are associated with the redder galaxies. As noted in Paper~2, the IEGs in the red end of the relationship may not have been detected in \HI\ due to removal of a large number of scans, reducing the net exposure times for those objects.

\begin{figure*}
\epsscale{0.9}
\plotone{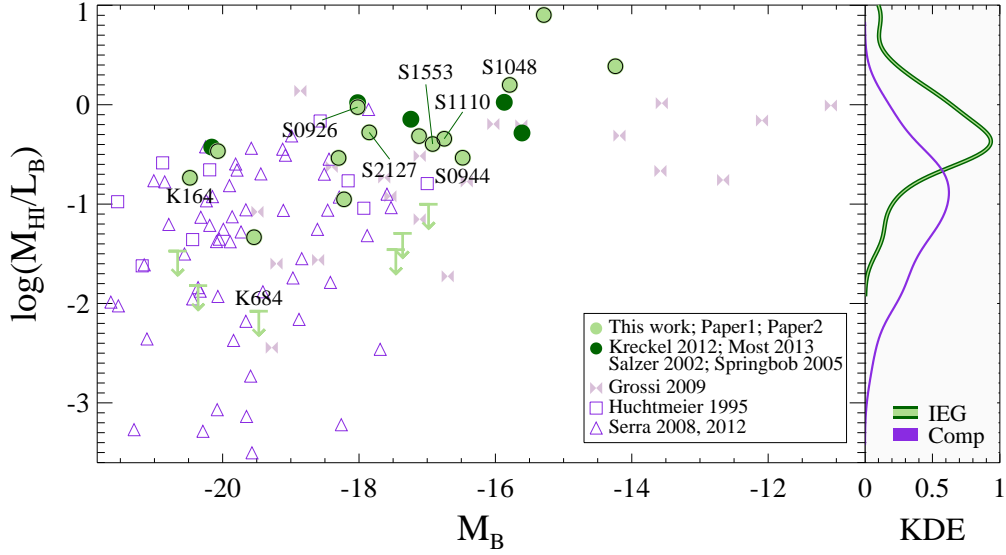}
\caption{{\bf Left panel:} The log of the \HI\ mass-to-blue luminosity ratio \frenchspacing{vs.} absolute blue magnitude for the IEG and comparison samples. The same color coding is used as in  Figure~\ref{fig:logMHILB_BV}. {\bf Right panel:} The kernel distribution estimation of the IEG (green) and comparison (purple) log(M$_{\rm{HI}}$/L$_{\rm{B}}$) distributions. This analysis includes only those galaxies having measured \HI\ fluxes ($\rm{S}_{\rm{HI}}\geq0.28~\rm{Jy}~\rm{km}~\rm{s}^{-1}$) that would have been detectable by our survey constraints. An offset between the IEG and comparison sample distributions is apparent.}
\label{fig:gas_to_mass}
\end{figure*}

Figure~\ref{fig:gas_to_mass} displays the relationship between the \HI\ gas mass to blue light ratio and the blue luminosity for the 25 IEG (green) and ETG comparison (purple) samples. The IEG sample adds data to the sparsely-populated low-luminosity end of this relation. The relationship indicates that lower luminosity galaxies are generally more gas-rich (higher M$_{\rm{HI}}$/L$_{B}$). While the IEGs fall almost completely within the locus of points defined by the comparison sample, the IEGs with detected \HI\ appear to hug the upper envelope.  A possible flattening in the upper envelope at $M_{\rm{B}} \lessapprox$ \n16.6 is suggested, the existence of which could be bolstered by additional low luminosity galaxy observations.

Kernel density estimations\footnote{using the {\tt density} function in the {\tt stats} v3.5.1 R package, with a Gaussian kernel and bandwidth set to {\tt bw.nrd0} (e.g., Silverman's ‘rule of thumb’).} (KDE)  of the log(M$_{\rm{HI}}$/L$_{\rm{B}}$) distributions for the IEG and comparison samples are also displayed in Figure~\ref{fig:gas_to_mass}. The distribution functions exclude upper limits and data for galaxies with \HI\ fluxes fainter than 0.28~Jy~km~s$^{-1}$, the smallest flux among our 14 \HI\ detections. This threshold \HI\ flux criterion removed one of the IEG galaxies, S1323 \citep{Most2013aj145_150}, from the analysis. The large offset between the two density functions imply a systematic \HI\ gas-richness enhancement ($\sim0.6$~dex in log(M$_{\rm{HI}}$/L$_{\rm{B}}$)) of IEGs relative to luminosity-matched galaxies in the comparison sample. 

\subsection{Statistical Analyses}
In order to quantify the statistical significance of the possible gas enhancement of the IEGs, a two-sample Anderson--Darling (A--D) test\footnote{using the {\tt ad.test} function in the {\tt kSample R package} \citet{Scholz2018CODE_ksamplesK}}  was applied to a subset of the data used for the density functions in Figure~\ref{fig:gas_to_mass}.  While the IEG \HI\ sample reflects a large range of physical distances (25-127~Mpc; see Equation~\ref{eqn:mass}) and blue luminosities ($0.7-234\times$10$^8$), the process of selecting targets having predicted \HI\ fluxes sufficiently high to obtain a 5$\sigma$ detection within $\sim$10 hour exposure may have inadvertently biased the IEG sample to higher  M$_{\rm{HI}}$/L$_{\rm{B}}$ values. One reassuring attribute of Figure~\ref{fig:gas_to_mass} that argues against the existence of such a bias is the fact that all five of the IEG observations conducted by other surveys also lay along the upper envelope in the log(M$_{\rm{HI}}$/L$_{\rm{B}}$) relation. Nevertheless, to insure that statistical comparisons between the samples are not affected by differing observational flux limits,  galaxies having \HI\ fluxes less than 0.28~Jy~km~s$^{-1}$ are excluded from the A--D test, reducing the IEG and comparison sample sizes to 18 and 74, respectively. The A--D analysis  rejects the null hypothesis that the IEG and the comparison sample are drawn from the same parent population with a p-value of $1\times10^{-6}$.

\begin{figure*}
\epsscale{0.9}
\plotone{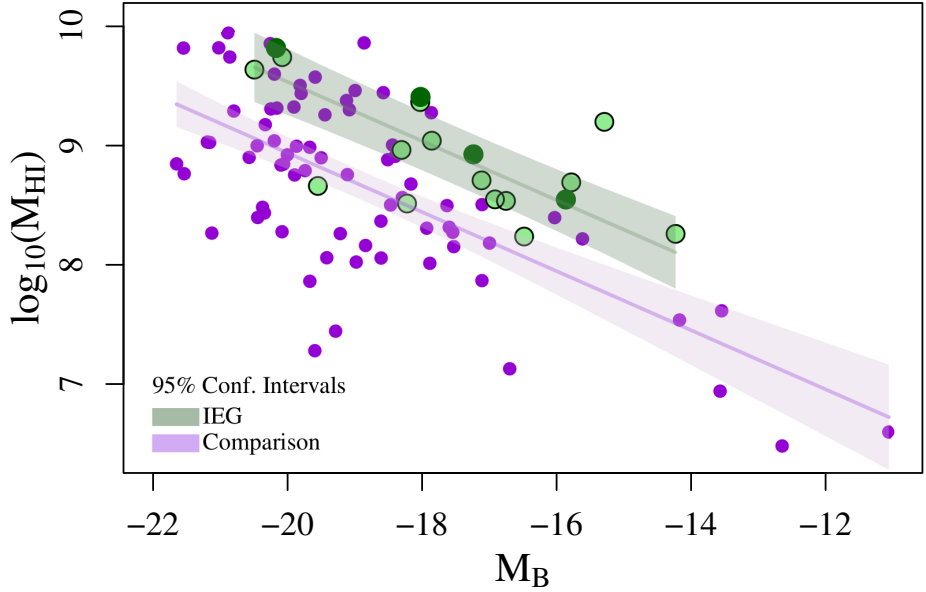}\\
\caption{Statistical analyses: The log of the \HI-mass \frenchspacing{vs.} absolute blue magnitude with same color legend as Figure~\ref{fig:logMHILB_BV} and displaying the same dataset ued to construct the KDEs displayed in Figure~\ref{fig:gas_to_mass}, e.g. those galaxies having $\rm{S}_{\rm{HI}}\geq0.28~\rm{Jy}~\rm{km}~\rm{s}^{-1}$ for statistical analysis having the same color legend as Figure~\ref{fig:gas_to_mass}. The fitted lines display the results of linear regression analysis; shaded regions are the 95\% confidence intervals for the IEG and comparison samples. The analysis supports the hypothesis that the \HI\ mass distributions of the two samples are distinct.\label{fig:95confint}}
\end{figure*}

The A--D test compares the two samples without regard for the relationship between M$_{\rm{HI}}$/L$_{\rm{B}}$ and M$_{\rm{B}}$. Figure~\ref{fig:gas_to_mass} suggests that lower gas-to-luminosity ratios are associated with galaxies of higher $B$ luminosities, the same end of the plot heavily represented by the comparison sample. To ensure that the A--D statistic is not at least in part driven by the comparison sample's M$_{\rm{B}}$ distribution being significantly weighted towards higher luminosities, multiple linear regression (MLR)\footnote{using the {\tt lm} function in the {\tt stats R package}} was applied to the same data subset used for the A--D test, shown in Figure~\ref{fig:95confint}.  

Applying standard techniques (see Appendix~\ref{sect:mlr}), a 1st-order MLR model adequately fits the data. In this model, log(M$_{\rm{HI}}$) is the dependent variable while M$_{\rm{B}}$ and the binary IEG/comparison sample identifier are the independent variables. The linear model is shown in Figure~\ref{fig:95confint} with the 95\% confidence intervals. 

The interaction between M$_{\rm{B}}$ and the sample indicator variable is not significant and it was removed from the model. The non-significance of the interaction term (p-value $=$ 0.4) indicates that the difference between the slopes of the fitted lines for the two samples is not statistically significant. Consequently, both slopes in Figure~\ref{fig:95confint} are equal. However, the sample indicator variable is significant with a p-value of $4\times10^{-5}$, which indicates that the vertical offset between the two lines is statistically significant. The coefficient for the indicator variable (0.6) represents the magnitude of this offset while controlling for M$_{\rm{B}}$. Because biases potentially exist when comparing the single-dish observations of the IEGs to the mixed single-dish and interferometric data of the comparison sample, MLR analysis was repeated on the subset of comparison data from only single-dish observations, with results indistinguishable from those described above. The statistical analyisis indicates that for a given luminosity, IEGs are significantly more gas enhanced than their counterparts in the comparison sample.

The MLR analysis supports the A--D test results that the $\sim0.6$~dex vertical (log(M$_{\rm{B}}$)) displacement between the two samples is statistically significant. However, the systematic \HI\ gas enhancement of the IEG sample appears to have statistical significance only when considered with respect to luminosity. When MLR analysis is applied using color (B$-$V) instead of M$_{\rm{B}}$ as an independent variable, the IEG and comparison samples are statistically indistinguishable in Figure~\ref{fig:logMHILB_BV}. In other words, the IEGs do not appear to be gas enhanced relative to comparison galaxies of similar optical color but are gas enriched when compared to galaxies of similar luminosity.

Other authors have found galaxies in loose groups, the field and cluster outskirts to have higher overall \HI\ mass-to-light ratios relative to ETGs in dense environments. Some have suggested this finding supports the emergence of early-type morphology via pre-processing inside galaxy groups \citep{Serra2012mnras422_1835} or of gas accretion from the intergalactic medium \citep{Grossi2009aap498_407, Keres2005mnras363_2, Maccio2006mnras366_1529}. The degree of \HI\ enhancement of ETGs in extremely isolated environments relative to field and loose group members is an unanticipated finding and one not previously explored, to our knowledge, in systems as isolated as the IEG sample.  The systematic \HI\ enhancement seen in the IEGs potentially provide a valuable clue in the understanding of ETG formation in low density environments, but exercising some caution in over-interpretation of the statistical analysis is prudent given the small IEG sample size.  If the enhancement is real, the origin of the excess gas is of interest. The gas could reflect an unusually low gas consumption rate in the galaxies' early history,  potentially a unique feature of extremely isolated systems.  The lack of enhancement in the comparison galaxies could be the result of neutral gas ``hidden" in a different gas phase or of \HI\ lost to interactions with higher density environments such as those of loose groups. 

\section{Conclusions}\label{sect:conclusions}

Seven of the eight IEGs presented in this study have detectable \HI\ emission.  Adding these systems to those presented in Papers~1 \& 2 and data from the literature for five other IEGs from other surveys, emission is detected in 19 of the 25 IEGs or about 76\% of the sample.  

Five of the galaxies presented as the targets of new observations in this paper have Gaussian-like \HI\ profiles.  \citet{Serra2012mnras422_1835} find that a majority of ETGs in low density environments with detectable \HI\ have regularly rotating gaseous disks or rings, suggesting that the detected \HI\ gas in these five galaxies is likely in disks which are either face-on or have near solid-body rotation and centrally-peaked radial gas density profiles characteristic of dwarf galaxies.  Over half (8 of the 14 detected IEGs) of the total \HI\ IEG sample presented here and in Papers 1 \& 2 have Gaussian-like \HI\ profiles. Of the other IEGs detected in \HI, 4 have two \HI\ peaks, 1 has a plateau or rectangular profile shape, and 1 has three \HI\ peaks.

We have compared the \HI\ gas-richness of our consolidated study of 25 IEGs to a comparison ETG sample drawn from \citet{Huchtmeier1995aap300_675}, \citet{Grossi2009aap498_407}, and \citet{Serra2012mnras422_1835, Serra2008aap483_57} consisting largely of loose group members.  Across both samples, bluer ETGs are generally more gas-rich. The IEG sample increases the sample coverage at the blue end of that relationship. We also find that the low luminosity galaxies (dwarfs) in the IEG sample are more gas-rich, consistent with the ETG comparison sample. Our study improves the coverage of this relationship at the sparsely populated low-luminosity end and finds that gas-richness is dramatically different between the comparison sample and the significantly more isolated IEGs. Across $\sim$2 orders of magnitude in luminosity, the IEGs are found to have, with a high level of statistical significance, gas masses that are systematically higher than those of galaxies of similar luminosity in the ETG comparison sample. Of particular note is that the IEG gas enrichment is seen not only in the blue ``dwarf''-like galaxies but also in the luminous red objects in the sample.

Other studies assessing the role of environment on the gas-richness of ETGs in the field, in loose groups and in the outskirts of clusters have concluded that such objects are generally more gas-rich than their cluster counterparts \citep{DiSeregoAlighieri2007aap474_851, Grossi2009aap498_407, Serra2012mnras422_1835}. The significant neutral gas enhancement of the isolated ETGs relative to the low density comparison sample may suggest that ETG mass assembly continues or that the gas of ETGs is retained in the absence of a galaxy group environment \citep{Serra2012mnras422_1835}. While some of the IEGs exhibit low surface brightness morphological evidence for past merging events in the optical bands pointing to a potential source of the \HI\ gas and highly asymmetric features found in their \HI\ line profiles, other systems may represent early-type galaxies that are continuing to assemble via quiescent \HI\ accretion from the cosmic web.     

\clearpage

\acknowledgments
T.A. is funded by an appointment to the NASA Postdoctoral Program at the NASA Ames Research Center, administered by Universities Space Research Association under contract with NASA. The Green Bank Observatory is a facility of the National Science Foundation operated under cooperative agreement by Associated Universities, Inc. This research has made use of the NASA/IPAC Extragalactic Database (NED), which is operated by the Jet Propulsion Laboratory, California Institute of Technology, under contract with the National Aeronautics and Space Administration.

\bibliographystyle{aasjournal}

\appendix
\setcounter{table}{0}
\renewcommand{\thetable}{A\arabic{table}}

\section{Standardization of Comparison Sample Data}\label{sect:compdata}
The ``standardized" data used for the comparison sample in this paper is presented in Table~\ref{tab:compinfo}. For details regarding the standardization process, see Section~\ref{sect:datastandardization}. 

\startlongtable
\begin{deluxetable*}{lllclrrrcc}
\tablecaption{Comparison Galaxy \HI\ and Optical Photometry\label{tab:compinfo}}
\tabletypesize{\footnotesize}
\tablecolumns{10}
\tablewidth{0pt}
\tablehead{
\colhead{Name} & \colhead{RA (2000)} & \colhead{Dec (2000)} &  \colhead{Distance} & \colhead{$\rm{M}_{\rm{B}}$} & \colhead{B$-$V} & \colhead{S$_{\rm{HI}}$} & \colhead{M$_{\rm{HI}}$} &  \colhead{log$_{10}$(M$_{\rm{HI}}$/L$_{\rm{B}}$)} & \colhead{Source}\\
\colhead{} & \colhead{(hh mm ss.s)} & \colhead{(dd mm ss)} & \colhead{(Mpc)} & \colhead{(mag)} &  \colhead{(mag)} & \colhead{(Jy \kms)} & \colhead{(10$^8$ M$_{\sun}$)} & \colhead{(M$_{\sun}$/L$_{\sun}$)} & \colhead{}\\
\colhead{(1)}  & \colhead{(2)} & \colhead{(3)}  & \colhead{(4)} & \colhead{(5)} &\colhead{(6)}  & \colhead{(7)}  & \colhead{(8)} & \colhead{(9)} & \colhead{(10)}}
\startdata
UGC1503 & 02 01 19.88 & +33 19 47.19 & 71 & \n20.2 & 0.82 & 3.30 & 39.5 & \n0.65 & 2\\
N807 & 02 04 55.66 & +28 59 14.79 & 66 & \n20.9 & 0.88 & 8.40 & 87.7 & \n0.58 & 2\\
N855 & 02 14 03.49 & +27 52 38.42 & 10 & \n17.0 & 0.65 & 6.20 & 1.5 & \n0.79 & 2\\
N1023 & 02 40 24.00 & +39 03 47.82 & 11 & \n20.2 & 0.95 & 67.06 & 20.5 & \n0.92 & 3\\
N1490 & 03 53 34.22 & \n66 01 05.01 & 70 & \n21.0 & \nodata & 5.70 & 65.5 & \n0.76 & 4\\
N1947 & 05 26 47.61 & \n63 45 36.10 & 12 & \n19.0 & 0.96 & 3.00 & 1.1 & \n1.74 & 4\\
UGC3960 & 07 40 22.74 & +23 16 29.98 & 32 & \n18.4 & 0.79 & 0.24 & 0.6 & \n1.79 & 3\\
N2534 & 08 12 54.15 & +55 40 19.43 & 50 & \n20.0 & 0.72 & 1.40 & 8.3 & \n1.25 & 4\\
N2594 & 08 27 17.16 & +25 52 43.73 & 34 & \n18.5\tablenotemark{a} & 0.84 & 2.80 & 7.6 & \n0.70 & 3\\
UGC4599 & 08 47 41.69 & +13 25 08.83 & 29 & \n18.9 & 0.82 & 35.34 & 72.1 & +0.14 & 1\\
N2685 & 08 55 34.71 & +58 44 03.83 & 16 & \n19.1 & 0.82 & 32.48 & 19.9 & \n0.51 & 3\\
N2764 & 09 08 17.46 & +21 26 36.02 & 38 & \n19.4 & 0.73 & 5.15 & 18.0 & \n0.70 & 3\\
N2768 & 09 11 37.50 & +60 02 14.00 & 23 & \n21.1 & 0.93 & 1.50 & 1.8 & \n2.36 & 4\\
N2824 & 09 19 02.23 & +26 16 11.93 & 40 & \n18.9 & 0.87 & 0.10 & 0.4 & \n2.16 & 3\\
N2810 & 09 22 04.49 & +71 50 38.46 & 53 & \n20.6 & 0.97 & 1.20 & 7.9 & \n1.50 & 4\\
N2859 & 09 24 18.53 & +34 30 48.55 & 26 & \n20.3 & 0.91 & 1.68 & 2.7 & \n1.88 & 3\\
Mrk706 & 09 34 03.03 & +11 00 21.73 & 35 & \n17.5 & 0.60 & 0.64 & 1.9 & \n0.92 & 1\\
SDSSJ093608.59+061525.4 & 09 36 08.59 & +06 15 25.42 & 34 & \n15.6\tablenotemark{a} & 0.54 & 0.60 & 1.6 & \n0.21 & 1\\
N2962 & 09 40 53.93 & +05 09 56.92 & 28 & \n19.5 & 0.98 & 4.22 & 7.9 & \n1.08 & 1\\
N2974 & 09 42 33.28 & \n03 41 56.90 & 26 & \n20.4 & 0.95 & 5.98 & 9.9 & \n1.36 & 2\\
N3032 & 09 52 08.15 & +29 14 10.36 & 25 & \n18.8 & 0.66 & 1.01 & 1.4 & \n1.55 & 3\\
CGCG064\n021 & 09 59 43.48 & +11 39 38.68 & 40 & \n18.6\tablenotemark{a} & 0.79 & 0.30 & 1.1 & \n1.56 & 1\\
N3073 & 10 00 52.07 & +55 37 07.79 & 20 & \n17.5 & 0.65 & 1.43 & 1.4 & \n1.04 & 3\\
N3108 & 10 02 29.02 & \n31 40 38.70 & 35 & \n20.2 & 0.96 & 6.90 & 20.2 & \n0.97 & 4\\
UGC5408 & 10 03 51.87 & +59 26 10.27 & 45 & \n18.5 & 0.56 & 0.67 & 3.2 & \n1.06 & 3\\
N3182 & 10 19 33.02 & +58 12 20.61 & 33 & \n19.7 & 0.94 & 0.03 & 7.9E\n02 & \n3.14 & 3\\
ESO092\n21 & 10 21 05.52 & \n66 29 31.40 & 25 & \n19.0 & \nodata & 19.20 & 28.8 & \n0.31 & 4\\
N3265 & 10 31 06.77 & +28 47 48.01 & 22 & \n17.9 & 0.73 & 1.77 & 2.0 & \n1.04 & 2\\
N3384 & 10 48 16.88 & +12 37 45.38 & 12 & \n19.6 & 0.91 & 0.59 & 0.2 & \n2.73 & 3\\
SDSSJ104926.70+121528.0 & 10 49 26.70 & +12 15 28.01 & 22 & \n14.2\tablenotemark{a} & 0.45 & 0.31 & 0.3 & \n0.31 & 1\\
N3414 & 10 51 16.20 & +27 58 30.36 & 24 & \n20.1 & 0.95 & 1.35 & 1.9 & \n1.93 & 3\\
SDSSJ105131.35+140653.2 & 10 51 31.34 & +14 06 53.17 & 12 & \n12.1\tablenotemark{a} & 0.69 & 0.22 & 7.1E\n02 & \n0.16 & 1\\
N3457 & 10 54 48.63 & +17 37 16.46 & 19 & \n17.9 & 0.77 & 1.23 & 1.0 & \n1.32 & 3\\
N3489 & 11 00 18.57 & +13 54 04.40 & 12 & \n19.3 & 0.82 & 0.86 & 0.3 & \n2.44 & 1\\
N3499 & 11 03 11.03 & +56 13 18.19 & 26 & \n17.7 & 0.94 & 0.04 & 6.2E\n02 & \n2.46 & 3\\
N3522 & 11 06 40.46 & +20 05 08.00 & 21 & \n17.6 & 0.74 & 1.92 & 2.1 & \n0.90 & 3\\
UGC6176 & 11 07 24.68 & +21 39 25.53 & 39 & \n18.4 & 0.92 & 2.76 & 10.1 & \n0.55 & 3\\
IC676 & 11 12 39.81 & +09 03 21.03 & 24 & \n19.2 & 0.77 & 1.33 & 1.8 & \n1.60 & 1\\
SDSSJ111445.02+123851.7 & 11 14 45.02 & +12 38 51.71 & 5 & \n11.1\tablenotemark{b} & 0.38 & 0.62 & 4.0E\n02 & \n0.01 & 1\\
N3608 & 11 16 58.95 & +18 08 55.26 & 22 & \n20.1 & 0.92 & 0.12 & 0.1 & \n3.07 & 3\\
IC2684 & 11 17 01.04 & +13 05 58.70 & 5 & \n12.7\tablenotemark{a} & 0.55 & 0.57 & 3.0E\n02 & \n0.76 & 1\\
SDSSJ111701.18+043944.2 & 11 17 01.17 & +04 39 44.21 & 24 & \n13.6\tablenotemark{a} & 0.44 & 0.30 & 0.4 & +0.02 & 1\\
N3619 & 11 19 21.55 & +57 45 28.16 & 26 & \n19.7 & 0.93 & 5.90 & 9.6 & \n1.06 & 3\\
N3626 & 11 20 03.81 & +18 21 24.62 & 26 & \n20.3 & 0.80 & 9.71 & 14.9 & \n1.13 & 3\\
IC692 & 11 25 53.47 & +09 59 14.95 & 19 & \n17.1 & 0.31 & 3.59 & 3.2 & \n0.52 & 1\\
N3773 & 11 38 12.87 & +12 06 43.37 & 10 & \n17.1 & 0.34 & 3.06 & 0.7 & \n1.15 & 1\\
IC719 & 11 40 18.50 & +09 00 35.59 & 29 & \n18.4 & 0.89 & 4.08 & 8.1 & \n0.63 & 1\\
UGC6655 & 11 41 50.64 & +15 58 25.50 & 5 & \n13.6 & 0.48 & 1.26 & 8.7E\n02 & \n0.67 & 1\\
2MASXJ11434609+1342273 & 11 43 46.11 & +13 42 27.26 & 42 & \n16.4\tablenotemark{a} & 1.08 & 0.22 & 0.9 & \n0.77 & 1\\
N3838 & 11 44 13.75 & +57 56 53.61 & 23 & \n18.6 & 0.90 & 1.84 & 2.3 & \n1.25 & 3\\
2MASXJ11460404+1134529 & 11 46 04.05 & +11 34 52.71 & 43 & \n17.6\tablenotemark{a} & 0.52 & 0.72 & 3.1 & \n0.73 & 1\\
N3928 & 11 51 47.62 & +48 40 59.28 & 18 & \n18.2 & 0.65 & 6.00 & 4.7 & \n0.77 & 2\\
N3941 & 11 52 55.36 & +36 59 10.78 & 16 & \n19.9 & 0.89 & 16.07 & 9.8 & \n1.13 & 3\\
N3945 & 11 53 13.72 & +60 40 32.00 & 23 & \n20.1 & 0.93 & 5.57 & 6.8 & \n1.38 & 3\\
N3998 & 11 57 56.13 & +55 27 12.92 & 19 & \n19.9 & 0.94 & 6.36 & 5.7 & \n1.38 & 3\\
N4026 & 11 59 25.19 & +50 57 42.09 & 18 & \n19.7 & 0.90 & 7.69 & 6.1 & \n1.28 & 3\\
N4036 & 12 01 26.75 & +61 53 44.80 & 24 & \n20.4 & 0.89 & 1.80 & 2.5 & \n1.95 & 3\\
N4111 & 12 07 03.13 & +43 03 56.59 & 14 & \n19.1 & 0.88 & 12.83 & 5.7 & \n1.06 & 3\\
N4125 & 12 08 06.02 & +65 10 26.90 & 24 & \n21.3 & 0.91 & 0.20 & 0.3 & \n3.27 & 4\\
N4150 & 12 10 33.65 & +30 24 05.49 & 13 & \n18.3 & 0.78 & 0.04 & 1.8E\n02 & \n3.22 & 3\\
N4203 & 12 15 05.05 & +33 11 50.37 & 20 & \n19.8 & 0.93 & 27.70 & 27.2 & \n0.66 & 3\\
N4278 & 12 20 06.82 & +29 16 50.72 & 16 & \n20.1 & 0.90 & 11.40 & 7.0 & \n1.35 & 4\\
N4521 & 12 32 47.65 & +63 56 21.13 & 39 & \n19.8 & 0.88 & 0.15 & 0.5 & \n2.37 & 3\\
ESO381\n47 & 13 01 05.39 & \n35 36 59.59 & 65 & \n20.2 & \nodata & 7.10 & 71.0 & \n0.42 & 4\\
IC4200 & 13 09 34.76 & \n51 58 06.89 & 52 & \n20.8 & \nodata & 8.70 & 54.9 & \n0.78 & 4\\
N5018 & 13 13 01.03 & \n19 31 05.49 & 40 & \n21.6 & 0.84 & 1.90 & 7.0 & \n1.99 & 4\\
N5103 & 13 20 30.07 & +43 05 02.30 & 23 & \n18.3 & 0.81 & 2.88 & 3.7 & \n0.93 & 3\\
N5173 & 13 28 25.27 & +46 35 29.93 & 38 & \n19.9 & 0.85 & 6.14 & 20.9 & \n0.81 & 3\\
N5198 & 13 30 11.40 & +46 40 14.79 & 39 & \n20.4 & 0.91 & 0.84 & 3.0 & \n1.84 & 3\\
N5338 & 13 53 26.55 & +05 12 27.95 & 10 & \n16.7 & 0.61 & 0.57 & 0.1 & \n1.73 & 1\\
N5422 & 14 00 42.03 & +55 09 52.10 & 30 & \n19.7 & 0.94 & 0.33 & 0.7 & \n2.18 & 3\\
N5557 & 14 18 25.72 & +36 29 36.81 & 48 & \n21.5 & 0.90 & 1.05 & 5.8 & \n2.02 & 3\\
N5582 & 14 20 43.12 & +39 41 36.90 & 25 & \n19.6 & 0.86 & 24.67 & 37.3 & \n0.44 & 3\\
N5631 & 14 26 33.29 & +56 34 57.46 & 32 & \n20.2 & 0.90 & 4.51 & 10.9 & \n1.21 & 3\\
SDSSJ144329.18+043153.4 & 14 43 29.18 & +04 31 53.41 & 28 & \n16.0\tablenotemark{a} & 0.50 & 1.36 & 2.5 & \n0.19 & 1\\
UGC9519 & 14 46 21.08 & +34 22 14.00 & 28 & \n17.9 & 0.98 & 10.36 & 18.8 & \n0.05 & 3\\
N5866 & 15 06 29.49 & +55 45 47.57 & 16 & \n20.3 & 0.84 & 0.17 & 0.1 & \n3.29 & 3\\
N5903 & 15 18 36.52 & \n24 04 06.89 & 36 & \n21.1 & 0.88 & 3.40 & 10.6 & \n1.61 & 4\\
ESO140\n31 & 18 37 53.62 & \n57 36 39.90 & 42 & \n19.8 & \nodata & 7.70 & 31.8 & \n0.60 & 4\\
N6798 & 19 24 03.17 & +53 37 29.20 & 37 & \n19.1 & \nodata & 7.23 & 23.7 & \n0.45 & 3\\
IC4889 & 19 45 15.14 & \n54 20 38.90 & 34 & \n20.8 & 0.90 & 7.10 & 19.4 & \n1.20 & 4\\
N7052 & 21 18 33.04 & +26 26 49.29 & 67 & \n21.2 & \nodata & 1.00 & 10.7 & \n1.62 & 2\\
N7280 & 22 26 27.58 & +16 08 53.59 & 28 & \n19.4 & 0.87 & 0.63 & 1.1 & \n1.88 & 3\\
N7332 & 22 37 24.54 & +23 47 53.99 & 20 & \n19.6 & 0.88 & 0.04 & 3.2E\n02 & \n3.50 & 3\\
N7426 & 22 56 02.85 & +36 21 40.90 & 76 & \n21.5 & \nodata & 4.80 & 65.3 & \n0.98 & 2\\
N7468 & 23 02 59.25 & +16 36 18.88 & 31 & \n18.6 & 0.39 & 12.50 & 27.6 & \n0.16 & 2\\
\enddata
\tablecomments{Column: (1) Designation taken from the Catalog of Isolated Galaxies \citep{Karachentseva1973soobschchspets8_3} or the Sloan Digital Sky Survey \citep{Abazajian2009apjs182_543}, respectively. (2)-(3) Galaxy coordinates sourced from the NASA/IPAC Extragalactic Database (NED). (4) Adopted distance (corrected for Virgo Infall). (5)-(6) Absolute B-band magnitude and B-V color, extinction corrected using \citet{Schlafly2011apj737_103}. (7) \HI\ flux density integrated over channels with detected emission. (8) Derived \HI\ mass. (9) \HI\ mass-to-blue luminosity ratio in solar units, using an absolute solar luminosity M$_{\sun, B} = 5.44$ \citep{Willmer2018apjs236_47}. (10) Data source: 1--\citet{Grossi2009aap498_407}; 2--\citet{Huchtmeier1995aap300_675}; 3--\citet{Serra2012mnras422_1835}; 4--\citet{Serra2008aap483_57}}
\tablenotetext{a}{Photometry derived through transformation of NSA {\tt Sersic flux} data; see Section~\ref{sect:datastandardization} for details.}
\tablenotetext{b}{Galaxy not in the NSA; photometry derived by transforming SDSS {\tt Cmodel} magnitudes listed in NED.}
\end{deluxetable*}

\clearpage
\section{Details of Statistical Analysis}\label{sect:mlr}
To determine whether the offset between the IEG and comparison samples in the M$_{\rm{B}}$ and log(M$_{\rm{HI}}$) (or equivalently log(M$_{\rm{HI}}$/L$_{\rm{B}}$)) relationship is statistically significant, various multiple linear regression models were fit to the data. The {\tt lm} function in the {\tt stats} v3.5.1 R package was used to perform the modeling. Independent modeling to confirm the results and interpretation was performed by one of us (J.D.F.) using {\tt Minitab} statistical software. In these models, M$_{\rm{B}}$ is the dependent variable. Independent variables are M$_{\rm{B}}$ and a binary indicator variable identifying the sample (e.g., 1 and 0 for IEG and comparison sample, respectively) associated with each observation. Interaction terms and a second-order term for M$_{\rm{B}}$ were tried. Equation~\ref{eqn:mlr} illustrates a general model: 

\begin{equation}\label{eqn:mlr}
log(M_{\rm{HI}}) = \beta_0 + \beta_1 M_{\rm{B}c} + \beta_2 S + \beta_3 S M_{\rm{B}c} + \beta_4 M_{\rm{B}c^2} + \beta_5 S M_{\rm{B}c^2} 
\end{equation}

where $\beta_0$ is the constant, M$_{\rm{B}c}$ is a ``centered'' variable, $\beta_i$ are coefficients, and $S$ identifies the sample. 

M$_{\rm{B}}$ was centered (mean value was subtracted) to reduce the multicollinearity that the interaction and 2nd-order terms would otherwise produce. Excessive levels of multicollinearity can hamper model selection by producing unstable coefficient estimates, reducing statistical power, and affecting p-values. Variance Inflation Factors (VIFs) were assessed to ensure that problematic amounts of multicollinearity were not present. VIFs were computed using the {\tt vif} function in the {\tt car R} library. VIFs less than $\sim$~5 indicate that multicollinearity is at an acceptable level. In the models for log(M$_{\rm{HI}}$), VIFs associated with the interaction and 2nd-order terms were found to have, at most, values less than 3.5. Additionally, confirmation of normally distributed residuals and of the assumption of homoscedasticity were performed by inspection of residuals, normal Q--Q, and scale-location graphical output of the {\tt R} {\tt lm} function.

The statistical output includes a p-value for each coefficient that allows one to assess the null hypothesis that the coefficient $\beta_i$ equals 0. The quadratic terms and interaction terms all had large ($\gtrapprox$0.1) p-values, including $\beta_3$ which quantifies the difference between the slopes of the fitted lines for the two samples. Following a standard model reduction process that systematically removes one non-significant term from the model at a time and then re-fits the data, the final model contains M$_{\rm{B}}$ and the binary sample indicator variable as the only significant independent variables (see Table~\ref{tab:mlr}).
 
The p-values for all three coefficients are very small. Most notable is the value associated with $\beta_2$, which is the vertical offset in the M$_{\rm{B}}$ versus log(M$_{\rm{HI}}$) relationship that exists between the IEG and comparison samples (see Figure~\ref{fig:95confint}). The corresponding p-value of $4\times10^{-5}$ suggests this offset of 0.6~dex is statistically significant.  

\begin{deluxetable}{cccc}[hb!]
\tablecaption{MLR Results For Final Model\label{tab:mlr}}
\tabletypesize{\footnotesize}
\tablecolumns{4}
\tablewidth{0pt}
\tablehead{
\colhead{} & \colhead{$\beta_0$} & \colhead{$\beta_1$} & \colhead{$\beta_2$}}
\startdata
Value      &  4.0                & -0.25               &   0.6\\
Std. Error &  0.5                & 0.03                &   0.1\\
p-value    &  $2\times10^{-11}$  & $2\times10^{-14}$   &   $6\times10^{-5}$\\
\enddata
\tablecomments{Because no interaction and quadratic terms are involved, this model uses M$_{\rm{B}}$ rather than the centered variable.}
\end{deluxetable}

\end{document}